\documentclass[twocolumn,aip,jcp,reprint,superscriptaddress,citeautoscript]{revtex4-1}

\usepackage[latin9]{inputenc}
\usepackage{amsmath}
\usepackage{amssymb}
\usepackage{graphicx}
\usepackage{esint}

\usepackage{enumerate}

\global\long\def\br{\mathbf r}
\global\long\def\rf{\rho_{f}}
\global\long\def\rb{\rho_{b}}
\global\long\def\sf{\sigma_{f}}
\global\long\def\sb{\sigma_{b}}
\global\long\def\upol{U_{\mathrm{pol}}}
\global\long\def\uelec{U_{\mathrm{elec}}}
\global\long\def\tuelec{{\tilde U}_{\mathrm{elec}}}
\global\long\def\cnum{\lambda_\mathrm{ratio}}
\global\long\def\mathd{\,\mathrm{d}}

\renewcommand{\epsilon}{\varepsilon}

\begin{document}

\title{Efficient and accurate simulation of dynamic dielectric objects}

\author{Kipton Barros}
\email{kbarros@lanl.gov}
\affiliation{Department of Materials Science and Engineering,
  Northwestern University, Evanston, Illinois 60208}
\affiliation{Department of Engineering Sciences and Applied Mathematics,
  Northwestern University, Evanston, Illinois 60208}
\affiliation{Theoretical Division and CNLS, Los Alamos National
  Laboratory, Los Alamos, New Mexico 87545}

\author{Daniel Sinkovits}
\affiliation{Department of Materials Science and Engineering,
  Northwestern University, Evanston, Illinois 60208}

\author{Erik Luijten}
\email{luijten@northwestern.edu}
\affiliation{Department of Materials Science and Engineering,
  Northwestern University, Evanston, Illinois 60208}
\affiliation{Department of Engineering Sciences and Applied Mathematics,
  Northwestern University, Evanston, Illinois 60208}

\begin{abstract}
  Electrostatic interactions between dielectric objects are complex and
  of a many-body nature, owing to induced surface bound charge. We
  present a collection of techniques to simulate \emph{dynamical}
  dielectric objects.  We calculate the surface bound charge from a
  matrix equation using the Generalized Minimal Residue method
  (GMRES). Empirically, we find that GMRES converges very
  quickly. Indeed, our detailed analysis suggests that the relevant
  matrix has a very compact spectrum for all non-degenerate
  dielectric geometries. Each GMRES iteration can be evaluated using a
  fast Ewald solver with cost that scales linearly or near-linearly in
  the number of surface charge elements.  We analyze several previously
  proposed methods for calculating the bound charge, and show that our
  approach compares favorably.
\end{abstract}
\maketitle

\section{Introduction}

Electrostatic interactions can induce complex behavior in
biological,\cite{perutz78,honig95,Clapham07}
colloidal,\cite{hunter01,leunissen05} and other\cite{Levin05,Vernizzi07}
soft-matter systems. Large-scale molecular dynamics and Monte Carlo
simulation of such mesoscale systems is only practical when the solvent
is treated as an implicit medium.  Moreover, the complexities associated
with induced polarization of dielectric media and the resulting
effective \emph{many-body} charge interactions are frequently ignored in
computational modeling.  We have developed an efficient method to
include complex dielectric interactions in the numerical investigation
of dynamical charge and dynamical (\emph{i.e.}, mobile) dielectric
media.  In Ref.~\onlinecite{barros13a}, we applied this method in the
first study of \emph{dynamical} colloids with dielectric many-body
interactions and observed surprising self-assembly phenomena.  Here, we
present a detailed account of the methodology.

Complex dielectric interactions arise because the dielectric medium
becomes electrically polarized in the presence of an applied electric
field~$\mathbf{E}$. The polarization field~$\mathbf{P}$ corresponds to a
local dipole density that partially cancels the applied field.  The
dielectric constant~$\kappa$ of a material controls the linear response
of polarization to the applied field (\emph{e.g.}, $\kappa \approx 2.6$
for polystyrene and $\kappa \approx 80$ for water at room
temperature). Within a uniform medium, polarization simply screens the
free charge, effectively reducing the electrostatic energy by a factor
$\kappa > 1$. The situation is far more interesting in regions where
$\kappa(\mathbf{r})$ varies, such as at interfaces between different
media. Here, the divergence of the polarization field gives rise to
bound charge that depends nonlocally on free charge sources, and
mediates effective interactions between charged objects.

Analytic solution of polarization charge and dielectric interactions is
limited to the simplest geometries. Dielectrophoresis has long been
studied~\cite{pohl51,pohl58}, but results are mostly limited to simple
dielectric objects~\cite{jones95}.  An implicit series expansion is
known for the system of two dielectric spheres~\cite{love75}.  For more
than two dielectric spheres, numerical treatment is
required~\cite{doerr04,doerr06}.  Ion dynamics in the presence of simple
dielectric geometries (\emph{e.g.}, a sphere or cylinder) can be solved
by specialized simulation techniques~\cite{messina02a,wynveen06}.
Explicit simulation of the solvent is, of course, also
possible~\cite{wynveen06}.  Several other bulk methods are
available. Clever Monte Carlo sampling of the full polarization field
allows generalization to nonlinear dielectric
media~\cite{maggs06,rottler09}. Alternatively, Car--Parrinello molecular
dynamics may be used to evolve the polarization
field~\cite{marchi01,levy05}.

In this paper, we analyze and extend an efficient method to simulate
electrostatic systems containing isotropic, linear dielectric media.
The electrostatic energy and forces follow directly from the bound
charge~$\rb(\mathbf{r})$, which we obtain by solving a matrix equation
involving the known free charge~$\rf(\mathbf{r})$ and dielectric
geometry~$\kappa(\mathbf{r})$. If the material boundaries are sharp,
$\rb(\mathbf{r})$ reduces to a \emph{surface} charge density
$\sb(\mathbf{r})$, which in turn greatly reduces the computational
cost. This general ``boundary-element'' approach to dielectrics has been
independently proposed several times, in multiple
forms~\cite{levitt78,zauhar85,hoshi87,allen01,boda04,jadhao12}. We
compare these methods, and argue that the surface bound
charge~$\sb(\mathbf{r})$ is most efficiently calculated using the
Generalized Minimum Residual (GMRES) method~\cite{Saad86}. Each GMRES
iteration requires a single matrix--vector product, which can be
calculated efficiently~\cite{Bharadwaj95,tyagi10} using a fast Ewald
(Coulomb) solver~\cite{karttunen08}. For example, the matrix--vector
product may be implemented with the Fast Multipole Method
(FMM)~\cite{Greengard87,greengard97}
or Lattice Gaussian Multigrid~\cite{sagui01} at a cost that scales
linearly in the number of discrete charge elements~$n$.

Empirically, we observe that GMRES converges rapidly to the solution
$\sb(\mathbf{r}) = x(\mathbf{r})$ of the matrix equation $\mathcal A x =
b$. This fast convergence may be attributed to the small condition
number of the linear operator~$\mathcal{A}$~\cite{liang97}.  We show
analytically that the eigenvalues of~$\mathcal{A}$ are bounded by
$\kappa_{\mathrm{min}} \leq \lambda \leq \kappa_{\mathrm{max}}$, the
extremal dielectric constants of the system.  We find that the ratio of
extreme eigenvalues, $\cnum = \lambda_\mathrm{max} /
\lambda_\mathrm{min}$ strongly controls GMRES convergence.  The
worst-case behavior, $\cnum = \kappa_{\mathrm{max}} /
\kappa_{\mathrm{min}}$ is realized in the dielectric slab
geometry. However, for ``typical'' geometries with non-degenerate aspect
ratios, we argue that $\cnum$ is of order unity, independent of the
dielectric constants $\kappa_{\mathrm{max}}$
and~$\kappa_{\mathrm{min}}$, \emph{provided} that we fix the net charge
on each dielectric object to its exact value (thereby eliminating an
outlying eigenvalue). By employing this and other optimizations, we find
that GMRES typically converges to order~$10^{-4}$ accuracy in only 3
or~4 iterations, each of which scales linearly in~$n$.  This high
efficiency has enabled our study of \emph{dynamical} dielectric
objects~\cite{barros13a}---to our knowledge, the first of its kind.

The remainder of this paper is organized as follows. In
Sec.~\ref{sec:review} we review the formulation of linear dielectrics as
a matrix equation to be solved for the surface bound charge. In
Sec.~\ref{sec:a_properties} we analytically bound the spectrum of the
relevant operator~$\mathcal{A}$, and argue that it is especially well
conditioned for typical dielectric geometries. In Sec.~\ref{sec:methods}
we discuss a collection of techniques that, in combination, enable
accurate and efficient simulation of dielectric systems. Finally, in
Sec.~\ref{sec:comparison} we analyze the convergence rates of several
recently proposed alternative methods, and argue that the combination of
GMRES with a fast Ewald solver is optimal.

\section{Review of linear dielectrics}
\label{sec:review}

\subsection{Electrostatic energy in a dielectric medium}

In the absence of a time-varying magnetic field, the electric field
satisfies~\cite{Landau93}
\begin{align}
  \nabla\cdot\mathbf{E} & = \rho/\epsilon_{0} \;, \label{eq:div_e} \\
  \nabla\times\mathbf{E} & = \mathbf{0} \;,
\end{align}
with $\rho(\br)$ the charge density field and $\epsilon_{0}$ the vacuum
permittivity.  The Helmholtz decomposition gives the electric field as
$\mathbf{E}=-\nabla\psi$, where the potential satisfies
$\nabla^{2}\psi=-\rho/\epsilon_{0}$.  It will be convenient to denote
the solution as $\psi=\mathcal{G}\rho/\epsilon_{0}$, where
\begin{equation}
  \mathcal{G}=-\nabla^{-2}
  \label{eq:gdef}
\end{equation}
is a linear operator. Its integral representation is
\begin{equation}
  (\mathcal{G}\rho)(\mathbf{r}) = \int_V
  G(\mathbf{r}-\mathbf{r}')\rho(\mathbf{r}')\mathd \br' \;,
\end{equation}
where the Green function $G(\mathbf{r})$ satisfies
$\nabla^{2}G(\mathbf{r})=-\delta(\mathbf{r})$.  If the system volume $V$
is infinite, $G(\mathbf{r})=1/4\pi|\mathbf{r}|$. Otherwise, we apply
periodic boundary conditions and Ewald summation.  The eigenvectors of
$\mathcal{G}$ are Fourier modes labeled by frequency~$\mathbf{k}$. Thus,
$\mathcal{G}$ commutes with derivatives, $\mathcal G \nabla = \nabla
\mathcal G$.  The eigenvalues of $\mathcal G$ are
$|\mathbf{k}|^{-2}$. We enforce charge neutrality to exclude the
$\mathbf{k}=\mathbf{0}$ mode, thus making $\mathcal{G}$ positive
definite. Finally, we note that $\mathcal{G}$ (like $\nabla^2$) is
symmetric, $\langle v, \mathcal G w \rangle = \langle \mathcal G v, w
\rangle$, under the inner product $\langle v, w \rangle = \int_V v(\br)
w(\br) \mathd \br$. This symmetry follows from the antisymmetry of
$\nabla$, which in turn follows from integration by parts (surface terms
do not appear, by the construction of $V$).

With this notation, the electric field becomes
\begin{equation}
  \mathbf{E} = -\nabla\mathcal{G}\rho/\epsilon_{0} \;.
  \label{eq:edef}
\end{equation}
In a dielectric medium, $\mathbf{E}$ will induce a polarization
(dipole-density) field~$\mathbf{P}$ with associated \emph{bound} charge
density,
\begin{equation}
  \nabla\cdot\mathbf{P}=-\rb \;.
  \label{eq:div_p}
\end{equation}
Thus, the total charge has both \emph{free} and bound components,
\begin{equation}
  \rho(\br)=\rf(\br)+\rb(\br) \;.
  \label{eq:rho_total}
\end{equation}
Moreover, the total (free) energy in a dielectric medium,
\begin{equation}
  U=\uelec+\upol \;,
  \label{eq:u}
\end{equation}
is a sum of the bare electric field energy,
\begin{equation}
  \uelec =
  \frac{\epsilon_{0}}{2}\int_V\mathbf{E}^{2} \mathd\br \;,
  \label{eq:uvac}
\end{equation}
and the free energy $\upol$ required to polarize
the medium~\cite{marcus56,felderhof77}.  Assuming an isotropic medium, a
Landau expansion in the polarization field yields, to lowest order,
\begin{equation}
  \upol =
  \frac{1}{2\epsilon_{0}} 
  \int_V\frac{\mathbf{P}^{2}}{\kappa-1}\mathd\br \;,
  \label{eq:upol}
\end{equation}
where $\kappa(\mathbf{r}) \geq 1$ is the dielectric constant of the
medium at position $\mathbf{r}$.  In equilibrium, $\mathbf{P}$ minimizes
$U$. Thus, we may solve $\delta U[\rf,\mathbf{P}]/\delta\mathbf{P}=0$ to
determine $\mathbf{P}$.  Beginning with Eq.~\eqref{eq:uvac}, we
substitute Eqs.~\eqref{eq:edef},~\eqref{eq:rho_total}
and~\eqref{eq:div_p}, and then apply the symmetry of $\mathcal{G}$ to
obtain
\begin{align}
  \frac{\delta \uelec}{\delta\mathbf{P}}
  & = \int_V\mathbf{E} \cdot \left(-\nabla\mathcal{G}
  \frac{\delta\rho}{\delta\mathbf{P}}\right) \mathd\br \nonumber \\
  & = \int_V\mathbf{E}\cdot\nabla\mathcal{G}\nabla\cdot
  \frac{\delta\mathbf{P}}{\delta\mathbf{P}}\mathd\br \nonumber \\
  & = \nabla\mathcal{G}(\nabla\cdot\mathbf{E}) \nonumber \\
  & = -\mathbf{E} \;.
\label{eq:duvac_dp}
\end{align}
Furthermore, Eq.~\eqref{eq:upol} implies
\begin{equation}
  \frac{\delta \upol}{\delta\mathbf{P}} = 
  \frac{\mathbf{P}}{\epsilon_{0}(\kappa-1)} \;,
  \label{eq:dupol_dp}
\end{equation}
so that the energy is minimized by a linear
polarization field,
\begin{equation}
  \mathbf{P}
  = \epsilon_{0}(\kappa-1)\mathbf{E} \;.
  \label{eq:p_propto_e}
\end{equation}
The quantity $\kappa(\mathbf{r}) - 1$ is the electric susceptibility of
the medium at position~$\br$.  Combination of
Eqs.~\eqref{eq:u},~\eqref{eq:uvac} and~\eqref{eq:upol} yields the total
energy,
\begin{equation}
  U=\frac{\epsilon_{0}}{2}\int_V\kappa\mathbf{E}^{2}\mathd\br \;.
  \label{eq:u_linear}
\end{equation}

Treatment of \emph{nonlinear} dielectric media is considerably more
difficult. Modifications to Eq.~\eqref{eq:upol} would yield a nonlinear
relation $\mathbf{E} = \delta \upol/\delta\mathbf{P}$, which
must be inverted to obtain $\mathbf{P}$ and
$\upol[\mathbf{P}]$.  The resulting energy will not be
quadratic in $\mathbf{E}$ and cannot be simply expressed as a sum of
pairwise charge interactions. A clever Monte Carlo approach to sample
$\mathbf{E}$ and~$\mathbf{P}$ in nonlinear dielectric media was proposed
in Ref.~\onlinecite{maggs06}.

\subsection{Bound charge formulation}

We now proceed to construct a linear operator equation for the bound
charge, and then formulate the electrostatic energy directly as a
function of free and bound charge.

We insert Eq.~\eqref{eq:p_propto_e} into Eq.~\eqref{eq:div_p},
and apply Eqs.~\eqref{eq:div_e} and~\eqref{eq:rho_total} to obtain
\begin{equation}
  \rb = -\nabla\cdot\mathbf{P} =
  -\epsilon_{0}\nabla\cdot\kappa\mathbf{E}+(\rf+\rb)
\end{equation}
and thus
\begin{equation}
  \epsilon_{0}\nabla\cdot\kappa\mathbf{E}=\rf \;.
  \label{eq:div_d}
\end{equation}
This equation is perhaps more familiar as $\nabla\cdot\mathbf{D}=\rf$,
with $\mathbf{D} = \epsilon_{0}\kappa\mathbf{E}$ the displacement field.
Substitution of Eq.~\eqref{eq:edef} then yields a fully explicit
relationship between free and bound charge,
\begin{equation}
  \mathcal{A}(\rf+\rb)=\rf \;,
  \label{eq:newconstitute}
\end{equation}
with
\begin{equation}
  \mathcal{A}=-\nabla\cdot\kappa\nabla\mathcal{G} \;.
  \label{eq:adef}
\end{equation}
The bound charge is the solution to the linear equation
\begin{equation}
  \mathcal{A}\rb = b \;,
  \label{eq:matrixeq}
\end{equation}
where the right-hand side is
\begin{equation}
  b = (1-\mathcal{A})\rf \;.
  \label{eq:bdef}
\end{equation}
When $b$, $\rf$, and~$\mathcal{A}$ are suitably discretized, one arrives
at a matrix equation equivalent to previous
works~\cite{levitt78,zauhar85,hoshi87,allen01,boda04}.

To formulate the energy as a function of charge, we substitute
Eq.~\eqref{eq:edef} into Eq.~\eqref{eq:u_linear} and integrate by parts,
\begin{equation}
  U = \frac{1}{2} \int_V (\nabla\cdot\kappa\mathbf{E})
  \mathcal{G}(\rf+\rb) \mathd\br \;. 
\end{equation}
Applying Eq.~\eqref{eq:div_d} we obtain
\begin{equation}
  U = \frac{1}{2\epsilon_{0}}
  \int_V\rf\mathcal{G}(\rf+\rb)\mathd\br = \frac{1}{2} \int_V \rf \psi \mathd\br \;,
  \label{eq:energybound}
\end{equation}
so that the energy follows immediately after solving
Eq.~\eqref{eq:newconstitute} for $\rb$.

\subsection{Charge screening}

A charged object in a uniform dielectric medium experiences screening
due to bound charge induced in the medium.  Consider a compact
domain~$\Omega$ enclosing an object so that there is uniform
dielectric constant~$\kappa_{\mathrm{bg}}$ on the
boundary~$\partial\Omega$. Applying the divergence theorem
yields
\begin{align}
  \int_{\Omega}\nabla\cdot\mathbf{E}\mathd\mathbf{r}
  & = \int_{\partial\Omega} \hat{n} \cdot \mathbf{E}\mathd\mathbf{s}
  = \kappa_{\mathrm{bg}}^{-1} \int_{\partial\Omega} \hat{n} \cdot
  \kappa\mathbf{E}\mathd\mathbf{s} \nonumber \\ 
  & = \kappa_{\mathrm{bg}}^{-1} \int_{\Omega} \nabla \cdot \kappa
  \mathbf{E}\mathd\mathbf{r} \;.
\end{align}
where, as usual, $\mathbf E (\mathbf r)$ and~$\kappa (\mathbf r)$ vary
with position $\mathbf r$.

Inserting Eqs.~\eqref{eq:div_e} and~\eqref{eq:div_d} gives the net
charge in the domain~$\Omega$,
\begin{equation}
  \int_{\Omega}(\rf+\rb)\mathd\mathbf{r} =
  \kappa_{\mathrm{bg}}^{-1}\int_{\Omega}\rf\mathd\mathbf{r} \;.
  \label{eq:netcharge} 
\end{equation}
This identity states that the net charge on a dielectric object is
a function only of its free charge and the surrounding dielectric
constant $\kappa_{\mathrm{bg}}$. Importantly, the dielectric constant
of the object itself does \emph{not} appear.

In regions where $\kappa(\mathbf{r})=\kappa_0$ is uniform, local
equality holds,
\begin{equation}
  \rf(\mathbf r)+\rb(\mathbf r)=\rf(\mathbf r)/\kappa_0 \;.
  \label{eq:netcharge_local} 
\end{equation}
This identity is also apparent from Eqs.\ \eqref{eq:newconstitute}
and~\eqref{eq:adef} when we set $\nabla\kappa=\mathbf 0$.

\subsection{Energy scaling}
\label{sub:scaling}

The following scaling argument provides intuition about when 
dielectric effects may be important.

A dielectric system is completely specified by the distribution of free
charge~$\rf(\br)$ and the geometry of the dielectric media
$\kappa(\br)$. If we scale
\begin{align}
  \rf(\br) & \rightarrow \alpha\rf(\br) \;, \\
  \kappa(\br) & \rightarrow \beta\kappa(\br) \;,
\end{align}
then by Eqs.~\eqref{eq:newconstitute} and~\eqref{eq:adef} the net
charge $\rho=\rf+\rb$ scales as $\rho\rightarrow(\alpha/\beta)\rho$.
By Eq.~\eqref{eq:energybound}, the energy scales as
\begin{equation}
  U\rightarrow(\alpha^{2}/\beta)U \;.
\end{equation}
Thus, the physics is invariant for any scaling that satisfies
$\alpha^{2}=\beta$.

Now consider a system of objects with dielectric
constant~$\kappa_{\mathrm{obj}}$ surrounded by a solvent with dielectric
constant~$\kappa_{\mathrm{bg}}$.  Choosing
$\alpha^{2}=\beta=1/\kappa_{\mathrm{bg}}$, we find that the system is
mathematically equivalent to one in which the objects have dielectric
constant
\begin{equation}
  \tilde{\kappa}=\kappa_{\mathrm{obj}}/\kappa_{\mathrm{bg}} \;,
  \label{eq:contrast}
\end{equation}
the solvent has dielectric constant~$1$, and all free charges are
divided by $\kappa_{\mathrm{bg}}^{1/2}$
[but note that the bound charge transforms in a more complicated way,
$\rb \rightarrow (\alpha / \beta) \rho - \alpha \rf = (\kappa_\mathrm{bg}^{1/2}
 - \kappa_\mathrm{bg}^{-1/2}) \rf + \kappa_\mathrm{bg}^{1/2} \rb$].
Thus, a single parameter $\tilde{\kappa}$ (the \emph{dielectric
  contrast}) controls the magnitude of dielectric effects.

Dielectric effects disappear when $\tilde \kappa = 1$; it is natural to
guess that they are maximized in the limits $\tilde{\kappa} \rightarrow
\{0, \infty \}$ of \emph{conducting} media (background or object,
respectively).  In Appendix~\ref{sec:energies} we plot the dielectric
energies of a point charge interacting with three prototypical
dielectric objects, namely a sphere, a cylinder, and a slab. For the
sphere, we find that the scaled dielectric energy effectively saturates
at $\tilde \kappa \approx 10^{\pm 1}$. We speculate that such saturation
is a universal feature of compact objects.  However, for extended
geometries such as the cylinder or the slab, the dielectric
energy may grow large in one (cylinder) or both (slab)
conducting limits.

\subsection{Reduction to surface charge}
\label{sub:surface}

Much numerical efficiency is gained by allowing $\kappa(\mathbf{r})$ to
vary only at sharp surface boundaries~\cite{levitt78}. We consider a
point $\mathbf{r}$ on a surface $S$ that separates regions of uniform
dielectric constant. The surface normal~$\hat{n}$ is defined to point
from $\kappa(\mathbf{r}) = \kappa_{\mathrm{in}}$
to~$\kappa_{\mathrm{out}}$.  Volume charge densities reduce to surface ones,
\begin{align}
  \rf(\mathbf{r}) & = \int_{S} \sf(\mathbf{r})
  \delta(\mathbf{r}-\mathbf{s}) \mathd\mathbf{s} \;,\\
  \rb(\mathbf{r}) & = \int_{S}\sb(\mathbf{r})\delta(\mathbf{r}-\mathbf{s})
  \mathd\mathbf{s} \;.
\end{align}
Our goal is to derive the counterpart of Eq.~\eqref{eq:matrixeq} for the
surface bound charge density~$\sb$.  Na\"{\i}ve application of
Eq.~\eqref{eq:div_d} presents difficult singularities at the
interface. To handle these singularities, we begin by integrating $\rf$
over an infinitesimal cylindrical (``pillbox'') volume~$\Omega$ that
encloses the surface point~$\mathbf{r}$. The cross-section of~$\Omega$ is a 
disk with area~$a$,
\begin{equation}
\int_{\Omega} \rf(\br') \mathd \br' = \sf(\br) \, a \;.
\end{equation}
Alternatively, Gauss's theorem applied to Eq.~\eqref{eq:div_d} gives
\begin{equation}
  \int_{\Omega} \rf(\br') \mathd \br' = \epsilon_{0}
  (\kappa_{\mathrm{out}}\mathbf{E}_{\mathrm{out}} -
  \kappa_{\mathrm{in}}\mathbf{E}_{\mathrm{in}})
  \cdot\hat{n} \, a \;,
\end{equation}
where $\mathbf{E}_{\mathrm{out/in}}(\mathbf{r}) =
\mathbf{E}(\mathbf{r}\pm\varepsilon\hat{n})$ for infinitesimal
$\varepsilon$. Thus
\begin{equation}
  \sf(\br) = \epsilon_{0} (\kappa_{\mathrm{out}}\mathbf{E}_{\mathrm{out}} -
  \kappa_{\mathrm{in}}\mathbf{E}_{\mathrm{in}}) \cdot\hat{n} \;. 
  \label{eq:surface_int_1}
\end{equation}
To relate $\mathbf{E}_{\mathrm{out/in}}$, we integrate the \emph{net}
charge density $\rho = \rf + \rb$ over the same pillbox volume
$\Omega$. This time, we apply Gauss's theorem to Eq.~\eqref{eq:div_e},
with the result
\begin{equation}
  \sf(\br) + \sb(\br) = \epsilon_0 (\mathbf E_\mathrm{out} - \mathbf
  E_\mathrm{in}) \cdot \hat n \;.
  \label{eq:surface_int_2}
\end{equation}
We wish to relate $\sf$ and $\sb$ via the average field
$\mathbf{E}(\mathbf{r}) =
(\mathbf{E}_{\mathrm{out}}+\mathbf{E}_{\mathrm{in}})/2$, which is
generated by external charges~$\rho(\mathbf{r}')$ for $\br' \neq
\br$. After some algebra, we obtain our desired result,
\begin{equation}
  \bar{\kappa}(\sf+\sb)+\epsilon_{0}\Delta\kappa\mathbf{E}\cdot\hat{n} =
  \sf \;,
  \label{eq:surface_linear}
\end{equation}
where
\begin{align}
  \bar{\kappa} & = (\kappa_{\mathrm{out}}+\kappa_{\mathrm{in}})/2 \;,\\
  \Delta\kappa & = \kappa_{\mathrm{out}}-\kappa_{\mathrm{in}}     \;.
\end{align}
It is interesting to compare Eq.~\eqref{eq:surface_linear} with the
volume-charge equivalent,
\begin{equation}
  \kappa(\rf+\rb)+\epsilon_{0}(\nabla\kappa)\cdot\mathbf{E}=\rf \;,
  \label{eq:volume_linear}
\end{equation}
obtained from na\"{\i}ve differentiation of Eq.~\eqref{eq:div_d} and
substitution of Eq.~\eqref{eq:div_e}. Since $\kappa(\mathbf{r})$ is
ill-defined \emph{at} a sharp dielectric boundary, reducing
Eq.~\eqref{eq:volume_linear} to Eq.~\eqref{eq:surface_linear} is
nontrivial.

We can write a linear equation for the surface bound charge analogous to
Eq.~\eqref{eq:matrixeq},
\begin{equation}
  \mathcal{A}\sb=b \;.
  \label{eq:matrixeq_surf}
\end{equation}
In this context, we replace Eqs.~\eqref{eq:adef} and~\eqref{eq:bdef}
with their surface-charge equivalents,
\begin{align}
  \mathcal{A}\sb & =
  \overline{\kappa}\sb+\epsilon_{0}\Delta\kappa\mathbf{E}_{b}\cdot\hat{n}
  \label{eq:adef_surf} \\
  b & =
  (1-\bar{\kappa})\sf-\epsilon_{0}\Delta\kappa\mathbf{E}_{f}\cdot\hat{n}
  \;.
  \label{eq:bdef_surf}
\end{align}
Here $\mathbf{E}_{b}(\mathbf{r}) = \int_{S}\sb(\mathbf{s})(\mathbf{r} -
\mathbf{s})/(4\pi\epsilon_{0}|\mathbf{r} - \mathbf{s}|^{3})
\mathd\mathbf{s}$ is the electric field due to surface bound
charge~$\sb$. To allow for the possibility of non-surface free charge,
we define $\mathbf{E}_{f}(\br) = \mathbf{E}(\br)-\mathbf{E}_{b}(\br)$ as
the electric field due to all charges \emph{other} than~$\sb$.

\subsection{Dielectric force}

The definition of force is conceptually straightforward: it is the
negative gradient of energy with respect to object position.  However,
evaluating this gradient for a dielectric object is somewhat subtle: One
must account for the complicated variation in~\emph{bound} charge as the
object moves~\cite{Landau93}.  In Appendix~\ref{sec:forces} we provide a
first-principles derivation of the total force on a rigid dielectric
object with fixed free charge,
\begin{align}
  \mathbf{F} & = \int_{\Omega}\mathbf{f}(\mathbf{r})\mathd\mathbf{r} \;,
  \label{eq:rigidforce}\\
  \mathbf{f}(\mathbf{r}) & = \kappa_{\mathrm{bg}}(\rf+\rb)\mathbf{E}
  \;,
  \label{eq:rigidforcedensity}
\end{align}
where $\Omega$ is a volume enclosing the object and its surface
charge. Torque on the rigid object is calculated in the natural way from
the force density~$\mathbf{f}(\mathbf{r})$. If the object has the same
dielectric constant as the background, $\kappa = \kappa_{\mathrm{bg}}$,
then the net charge is $\rf+\rb = \rf/\kappa_{\mathrm{bg}}$
[Eq.~\eqref{eq:netcharge_local}], and $\mathbf{F}$ reduces to the
standard Coulomb force.

Equation~\eqref{eq:rigidforce} may be understood physically by the
principle of effective moments~\cite{jones95}. We construct a
\emph{virtual} system in which the dielectric object under consideration
is replaced by a virtual object with a dielectric
constant~$\kappa_{\mathrm{bg}}$ that matches the background. The net
charge density $\rho = \rf + \rb$ on the physical and on the virtual
object is kept equal. Thus, by Eq.~\eqref{eq:edef}, the electric field
is also the same for the physical and the virtual system. The principle
of effective moments then states that the force on the physical and on
the virtual object is equal. In the virtual system, $\kappa(\br) =
\kappa_{\mathrm{bg}}$ is uniform, and the usual Coulomb force expression
applies, $\mathbf F =
\int_{\Omega}\tilde{\rho}_f\mathbf{E}\mathd\mathbf{r}$. Note that the
virtual free charge~$\tilde{\rho}_f$ differs from the physical free
charge~$\rf$. After accounting for dielectric screening in the virtual
system, Eq.~\eqref{eq:netcharge_local}, we obtain
$\tilde{\rho}_f/\kappa_{\mathrm{bg}} = \rf+\rb$. Combining the above
results, we reproduce Eq.~\eqref{eq:rigidforce}.

\subsection{Dielectric stress tensor}

The standard formula for virial stress also applies to a collection of
dielectric objects~\cite{sinkovits13},
\begin{equation}
  \tau = -
  \frac{1}{|\Omega|}\sum_{k}
  \left[m_{k}\mathbf{v}_{k}\otimes\mathbf{v}_{k}+\frac{1}{2}
    \sum_{\ell}\mathbf{r}_{k\ell}\otimes\mathbf{F}_{k\ell}\right] \;,
\end{equation}
where $\mathbf{r}_{k\ell}=\mathbf{r}_{k}-\mathbf{r}_{\ell}$ is the
displacement vector between the objects' centers of mass, and
$\mathbf{F}_{k\ell}$ is the force applied \emph{on} dielectric
object~$k$ \emph{by} the field~$\mathbf{E}$ generated by object~$\ell$
[Eqs.~\eqref{eq:rigidforce} and~\eqref{eq:rigidforcedensity}]. For
periodic boundary conditions, the sum over~$\ell$ should be extended to
include all periodic images. To address a potential source of confusion:
Although dielectric interactions are many-body in nature, we are using
the fact that, once the bound charge is known, forces and energies can
be expressed pairwise~\cite{sinkovits13}.

\section{Properties of the operator $\mathcal{A}$}
\label{sec:a_properties}

The efficient numerical solution of Eq.~\eqref{eq:matrixeq} depends on
the properties of operator $\mathcal{A}$, Eq.~\eqref{eq:adef}. We
demonstrate that $\mathcal{A}$ is diagonalizable and that its
eigenvalues are real and bounded by the extremal dielectric constants
contained in the system. Our results characterize the action of
$\mathcal{A}$ for any free charge density. In particular, they remain
valid in the limiting case of a \emph{surface} charge density, in which
the action of $\mathcal A$ is given by Eq.~\eqref{eq:adef_surf}.

The operator $\mathcal{A}=-\nabla\cdot\kappa\nabla\mathcal{G}$ is not
symmetric because its (symmetric) factors, $\nabla\cdot\kappa\nabla$ and
$\mathcal{G}$, do not generally commute when $\kappa(\mathbf{r})$ is
spatially varying. Similarly, $\mathcal{A}$ is not normal
($\mathcal{A}\mathcal{A}^{T}\neq\mathcal{A}^{T}\mathcal{A}$) and is not
expected to have an orthogonal eigenbasis. However, $\mathcal{G}$ is
symmetric and positive definite so we \emph{can} diagonalize the
symmetric operator,
\begin{equation}
\begin{split}
  \mathcal{G}^{1/2}\mathcal{A}\mathcal{G}^{-1/2} & =
  -\mathcal{G}^{1/2}(\nabla\cdot\kappa\nabla)\mathcal{G}^{1/2}\\
  & =\mathcal U\Lambda \mathcal U^{-1} \;,
\end{split}
\end{equation}
with unitary $\mathcal U$. Thus, $\mathcal{A}$ can be diagonalized, 
\begin{equation}
  \mathcal{A}=(\mathcal{G}^{-1/2}\mathcal U)\Lambda(\mathcal{G}^{-1/2}\mathcal U)^{-1} \;.
\end{equation}

An arbitrary eigenvector~$v$ of $\mathcal{A}$, with corresponding
eigenvalue~$\lambda$, satisfies
\begin{align}
  0 &=\mathcal{A}v-\lambda v \nonumber \\
  &=(-\nabla\cdot\kappa\nabla\mathcal{G}-\lambda)v \nonumber \\
  &=-\nabla\cdot(\kappa-\lambda)\nabla\mathcal{G}v \;,
\end{align}
where we have made use of the identity $\nabla^{2}\mathcal{G}=-1$.
We take the inner product of this equation with the vector $\mathcal{G}v$
and integrate by parts to get 
\begin{equation}
  0 = \int_V (\kappa-\lambda)|\nabla\mathcal{G}v|^{2} \mathd\br \;. 
  \label{eq:eigencond}
\end{equation}
This equation bounds the eigenvalues. If, for example, $\lambda$ were
greater than~$\kappa_{\max}$, the maximum value of~$\kappa(\mathbf{r})$
in the domain, the right-hand side would assuredly be negative,
violating the equality. The conclusion is that
\begin{equation}
  1<\kappa_{\min} \leq \lambda \leq \kappa_{\max} \;,
  \label{eq:eigenbounds}
\end{equation}
where the left-most bound is physical.

The condition number $||\mathcal A|| \cdot ||\mathcal A^{-1}||$ of
$\mathcal A$ is a good indicator of the difficulty of solving the
discretized matrix equation $\mathcal A x = b$. In particular, the
condition number measures the sensitivity of~$x$ to perturbations
in~$b$.  A closely related quantity is the ratio of extremal
eigenvalues,
\begin{equation}
  \cnum = \frac{\max_\lambda |\lambda|}{\min_\lambda |\lambda|} \;.
  \label{eq:lambda_ratio}
\end{equation}
Indeed, the condition number would be exactly $\cnum$ if $\mathcal A$
were normal. In Sec.~\ref{sub:convergence} we will observe that the
GMRES convergence rate is strongly linked to $\cnum$.

From Eq.~\eqref{eq:eigenbounds}, we see that $\cnum \leq
\kappa_\mathrm{max} / \kappa_\mathrm{min}$. In
Appendix~\ref{sec:exact_spectra} we solve the exact spectra for
sphere [Eq.~\eqref{eq:lambda_sphere}], cylinder
[Eq.~\eqref{eq:lambda_cyl}], and slab [Eq.~\eqref{eq:lambda_slab}]
geometries. After eliminating the constant eigenvector by fixing the net
object charge, Eq.~\eqref{eq:netcharge}, we find the following:
\begin{enumerate}
\item For the sphere, $\cnum \leq 3$, regardless of
  $\kappa_{\mathrm{out}}$ and~$\kappa_{\mathrm{in}}$.
\item For the infinite cylinder, $\cnum$ is small, except if
  $\kappa_{\mathrm{in}}\gg\kappa_{\mathrm{out}}$, in which case $\cnum
  \approx\kappa_{\mathrm{in}}/\kappa_{\mathrm{out}}$.  However, if the
  ratio of length~$L$ to radius~$R$ is not too big, then $\cnum$ is
  always small. For $L/R=30$, we estimate $\cnum \approx 13$, even when
  $\kappa_{\mathrm{in}}/\kappa_{\mathrm{out}} \rightarrow \infty$.
\item The infinite slab is the worst-case geometry, and demonstrates that
  the bounds of Eq.~\eqref{eq:eigenbounds} are tight. However, as in the
  cylindrical case, we expect better behavior when the slab has finite
  extent.
\end{enumerate}
These exact results suggest that for compact geometries (\emph{i.e.},
those with finite aspect ratio) the eigenvalue ratio~$\cnum$ will be
order unity, independent of~$\kappa_{\max}/\kappa_{\min}$.

In Appendix~\ref{sec:energies} we plot the exact energies of the sphere,
cylinder, and slab as a function of dielectric
contrast~$\tilde{\kappa}$.  We find that when $\cnum$ is small, the
energies saturate quickly as a function of dielectric contrast.
Conversely, large $\cnum$ implies stronger dielectric effects due
to greater accumulation of bound charge associated with long-wavelength
eigenvectors of~$\mathcal A$.

\section{Implementation Details and Considerations}
\label{sec:methods}

In a numerical study, it is convenient to calculate the energy via
Eq.~\eqref{eq:energybound}, \emph{i.e.}, in terms of free and bound
charge. The bound charge may be calculated by solving
Eq.~\eqref{eq:matrixeq}.  In dielectric geometries with sharp surface
boundaries, we instead solve Eq.~\eqref{eq:matrixeq_surf} for the
\emph{surface} bound charge~$\sb$. In this section, we discuss how to
discretize this linear equation for $\sb$ and solve it iteratively by
the Generalized Minimum Residual (GMRES) method~\cite{Saad86}. Each
iteration of GMRES requires only a single matrix--vector product, which
can be evaluated efficiently with a fast Ewald solver to solve the
vacuum electrostatic problem (several such routines are reviewed in
Ref.~\onlinecite{karttunen08}).  The surface bound charge may readily be
used to calculate both energy and forces.  Thus, our method is suitable
to the molecular dynamics simulation of mobile dielectric objects.  The
total computational cost per time step is then $\mathcal{O}(n)$ or
$\mathcal{O}(n\ln n)$, depending on the Ewald solver, where $n$ is the
number of surface patch elements.

\subsection{Discretization}
\label{sub:Discretization}

Numerical evaluation of Eq.~\eqref{eq:matrixeq_surf} requires
discretization of the surface into patch elements. Each surface patch~$i$
has a position~$\mathbf{r}_i$, a normal vector~$\hat{n}_i$, and a surface
area~$a_i$. The matrix--vector product~$\mathcal{A}\sigma$ is discretized
as $\sum_{j}\mathcal{A}_{ij}\sigma_{j}$, where
\begin{equation}
  \mathcal{A}_{ij} =
  \overline{\kappa}_{i}\delta_{ij}+\Delta\kappa_{i}\hat{n}_{i}
  \cdot \mathbf{I}_{ij}a_{j} \;,
  \label{eq:matrixop}
\end{equation}
and $\mathbf{I}_{ij}a_{j}\sigma_{j}/\epsilon_{0}$ is the electric field
on the $i^{\mathrm{th}}$ patch due to the surface charge at the
$j^{\mathrm{th}}$ patch. The vector~$b_{i}$ is similarly discretized. In
an infinite system, for example, we take the interaction elements to be
\begin{equation}
  \mathbf{I}_{ij} = (\mathbf{r}_{i} - \mathbf{r}_{j}) /
  4\pi|\mathbf{r}_{i} - \mathbf{r}_{j}|^{3} \;.
  \label{eq:pointinteraction}
\end{equation}
With periodic boundary conditions, Ewald summation should be used
instead.

\subsection{Patch corrections}
\label{sub:patch_corr}

As written in Eq.~\eqref{eq:pointinteraction},
$\mathbf{I}_{ii}$ exhibits an unphysical divergence. To lowest
order, one may assume the self-interactions to be zero,
$\mathbf{I}_{ii}=\mathbf{0}$. We obtain a better approximation to the self-field
by averaging contributions over the entire patch surface~$S_{i}$ with
area~$a_i$ and center point~$\mathbf{r}_{i}$,
\begin{equation}
  \mathbf{I}_{ii} = \frac{1}{a_i} \int_{S_{i}}
  \frac{\mathbf{r}_{i}-\mathbf{s}}
  {4\pi |\mathbf{r}_{i}-\mathbf{s}|^{3}} \mathd\mathbf{s} \;.
\end{equation}
If we assume that $S_{i}$ is disk shaped with area~$a_{i}$ and mean
curvature $\varkappa_{i}\ll a_{i}^{-1/2}$ then, after a lengthy calculation, we obtain
\begin{equation}
  \mathbf{I}_{ii}
  = \frac{\varkappa_{i}\hat{n}_{i}}{4 \sqrt{\pi a_{i}}} \;.
  \label{eq:curvaturecorr}
\end{equation}
In practice, this approximation works reasonably well for arbitrary
patch geometry and generalizes previous results for cylinder and sphere
patches~\cite{hoyles98,allen01}. The self-interaction in
Eq.~\eqref{eq:curvaturecorr} contributes to $\mathcal{A}$ at order
$\mathbf{I}_{ii}a_{i}\sim\sqrt{a_{i}}$.  Since this correction
is only approximate, we expect errors at the same order.

This type of correction may be generalized to interactions between
distinct patches.  For example, Eq.~\eqref{eq:pointinteraction} may be
replaced with an integral,
\begin{equation}
  \mathbf{I}_{ij} = \frac{1}{a_j} \int_{S_{j}}
  \frac{\mathbf{r}_{i}-\mathbf{s}}%
  {4\pi|\mathbf{r}_{i}-\mathbf{s}|^{3}}\mathd\mathbf{s}
  \;.
  \label{eq:patch-patch}
\end{equation}
Such treatment is primarily useful for nearby patches.  When similar
surface integrals are also applied to the energy calculation, the scheme
is called SC/SC in Ref.~\onlinecite{boda05}.  Higher-order corrections
are also possible. A natural next step is to replace
Eq.~\eqref{eq:patch-patch} with a double integral over both surface
patches~\cite{Tausch01,bardhan09}.  Full numerical evaluation of these
integrals is most practical for static dielectric geometries, or within
a rigid dielectric object, where the matrix elements $\mathcal{A}_{ij}$
are fixed.

In dynamic geometries, large discretization errors may occur in regions where a point
charge approaches a dielectric surface, or where two dielectric surfaces
approach each other. To improve accuracy in such cases, a natural
strategy is adaptive mesh refinement, in which patches are
recursively subdivided until some threshold is met.  For example, one
may require that the distance between neighboring patches should be some
factor less than the distance between the surface and the external
charge.

\subsection{GMRES}

The generalized minimum residual (GMRES) method solves $\mathcal{A}x=b$,
yielding $\sb=x$ by Eq.~\eqref{eq:matrixeq_surf}, without explicitly
constructing $\mathcal{A}^{-1}$. At the $m^{\mathrm{th}}$ iteration,
GMRES builds the Krylov space,
\begin{equation}
  K^{(m)}=\mathrm{span}\{b,\mathcal{A}b,\ldots,\mathcal{A}^{m-1}b\} \;.
\end{equation}
From within this space, GMRES selects the \emph{optimal} approximation
$x^{(m)}\in K^{(m)}$ to $x$, in the sense that $x^{(m)}$ minimizes the
norm $||r^{(m)}||=\sqrt{\langle r^{(m)}, r^{(m)}\rangle}$ of the
residual
\begin{equation}
  r^{(m)}=b-\mathcal{A}x^{(m)} \;.
\end{equation}
Here, the natural inner product is the discretized surface integral,
$\langle x,y\rangle=\sum_{i}x_{i}y_{i}a_{i}$, where $a_{i}$ is the area
of the $i^{\mathrm{th}}$ patch.

At the $m^{\mathrm{th}}$ GMRES iteration, the $m$-dimensional vector
space $K^{(m)}$ must be orthogonalized, at a cost that scales as
$\mathcal{O}(mn)$, because each vector contains $n$ surface patches. In
practice, GMRES converges in so few iterations (cf.\
Sec.~\ref{sub:convergence}) that the cost of orthogonalization is
negligible compared to the cost of building $K^{(m)}$. In particular,
``restarting'' GMRES is unnecessary.

\subsection{Fast matrix--vector product}
\label{sec:matrix_vec}

The dominant cost of GMRES is evaluating the matrix--vector products
needed to build the Krylov space. Referring to Eqs.~\eqref{eq:adef_surf}
and~\eqref{eq:matrixop}, we find that the key task is to calculate the
electric field~$\mathbf E_b$ generated by~$x^{(m)}$ (the $m$th iterative
approximation to~$\sigma_b$) and evaluated at every surface patch.  A
na\"{\i}ve implementation requires summing over all $\mathcal{O}(n^{2})$
pairs of patches. A fast Ewald solver such as particle--particle
particle--mesh (PPPM)~\cite{hockney-eastwood-first,Pollock96}, smooth
particle--mesh Ewald (PME)~\cite{essmann95}, or lattice gaussian
multigrid (LGM)~\cite{sagui01} reduces the cost to $\mathcal{O}(n\ln n)$
(for PPPM and PME) or $\mathcal{O}(n)$ (for LGM), \emph{provided} that
the charges are distributed uniformly in the system volume. The fast
multipole method (FMM), which costs $\mathcal{O}(n)$~\cite{Greengard87,greengard97},
may be better suited to the non-uniform distributions typical of surface
patches. These and other fast Ewald solvers are reviewed in
Ref.~\onlinecite{karttunen08}. In our implementation, we employed the
PPPM routine provided by LAMMPS~\cite{plimpton95}.

\subsection{Convergence criterion}

At every iteration, GMRES constructs the vector $x^{(m)}$ in the Krylov
space that minimizes the norm of the residual $||r^{(m)}||$. Although it
is not guaranteed, empirically we find that the relative errors in the
bound charge, $||x-x^{(m)}|| / ||x||$, and in the energy,
$|U(x)-U(x^{(m)})| / |U(x)|$, both have approximate magnitude
$||r^{(m)}|| / ||b||$. With the condition
\begin{equation}
  ||r^{(m)}|| < 10^{-4}||b||
  \label{eq:conv_criterion}
\end{equation}
we observe that the relative error in the energy is $\approx 10^{-4}$.

\subsection{Convergence rate}
\label{sub:convergence}

\begin{figure}
  \includegraphics{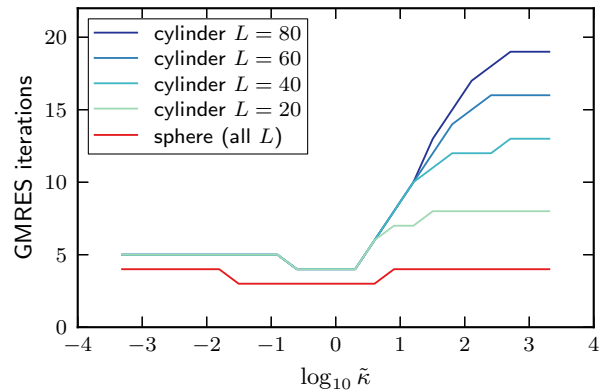}
  \caption{Number of GMRES iterations required to calculate the
    polarization charge~$\sb$ on a dielectric cylinder or sphere as a
    function of the dielectric contrast~$\tilde{\kappa}$. We use cylinders
    with unit radius and different lengths~$L$. Convergence is generally
    reached very quickly, except in geometries with extreme aspect
    ratios \emph{and} extreme~$\tilde{\kappa}$. The convergence
    threshold is $||r^{(m)}||<10^{-4}||b||$
    [Eq.~\eqref{eq:conv_criterion}], where $r^{(m)}=b-\mathcal{A}
    x^{(m)}$ is the residual of the $m^{\mathrm{th}}$ iterative
    approximation~$x^{(m)}$ to the bound charge~$\sigma_b$. We observe
    empirically that our convergence threshold corresponds to a relative
    error in the electrostatic energy of approximately $10^{-4}$.}
  \label{fig:convergence}
\end{figure}

In practice, we observe that GMRES finds the bound charge in very few
iterations. This observation is supported by mathematical properties of
the GMRES algorithm~\cite{Saad86}.  Because $\mathcal{A}$ is positive
definite [cf.\ Eq.~\eqref{eq:eigenbounds}], the residual error decreases
exponentially with the number of iterations.  If $\mathcal{A}$ were also
symmetric, then its condition number could be used to bound the rate of
GMRES convergence.  For our non-symmetric operator, less is known
analytically. Here, we demonstrate empirically that the convergence rate
is linked to the ratio of extremal values, $\cnum = \lambda_\mathrm{max}
/ \lambda_\mathrm{min}$ [Eq.~\eqref{eq:lambda_ratio}].  In
Sec.~\ref{sec:a_properties} we bounded $\cnum \leq
\kappa_{\mathrm{max}}/\kappa_{\mathrm{min}}$, and estimated $\cnum$ for
sphere, cylinder and slab geometries.

Figure~\ref{fig:convergence} demonstrates the link between fast GMRES
convergence and the smallness of $\cnum$.  For a single sphere $\cnum <
3$, and GMRES converges within a handful of iterations regardless of the
dielectric contrast
$\tilde{\kappa}=\kappa_{\mathrm{obj}}/\kappa_{\mathrm{bg}}$.  In
molecular dynamics simulations of many spheres in various
configurations, we observed GMRES convergence almost identical to that
of the single-sphere system~\cite{barros13a}.  The worst-case
convergence rates occur in geometries with extreme aspect ratios.  For a
cylinder, we predict $\cnum \approx
\kappa_{\mathrm{max}}/\kappa_{\mathrm{min}}$ only when
$\tilde{\kappa}\gg1$ and the cylinder length~$L$ is much larger than its
radius.  Indeed, this is precisely the regime where
Fig.~\ref{fig:convergence} shows slowed GMRES convergence. We observe
that, at fixed accuracy, the number of GMRES iterations scales like $\ln
\cnum$.

\subsection{Treatment of isolated point charges}

We allow systems to contain isolated point charges in addition to
dielectric objects, a situation that typically occurs in simulations
involving ionic solutions. In the bulk of a medium, where the dielectric
constant $\kappa(\br) = \kappa_0$ is uniform,
Eq.~\eqref{eq:netcharge_local} states that free charges are screened by
the factor~$\kappa^{-1}_0$. In numerics we typically deal with isolated
free charges~$q_f$, to which we must associate a net (free and bound) charge $q_f /
\kappa_0$.

Thus, only the surface bound charge~$\sb$ on the dielectric objects
remains to be calculated. To do so, we solve Eq.~\eqref{eq:adef_surf}
with $b$ in Eq.~\eqref{eq:bdef_surf} defined via the electric
field~$\mathbf{E}_{f}$ generated by both free surface charge~$\sf$ and
screened point charges~$q_{f}/\kappa_0$.

\subsection{Fixing net charge on objects}
\label{sub:fix_net_charge}

By Eq.~\eqref{eq:netcharge}, we may also fix the net integrated charge
on dielectric \emph{objects}. In particular, if the object is surrounded
by a medium with uniform dielectric constant~$\kappa_{\mathrm{bg}}$ and
carries total free charge~$q$ (counting both internal and surface
charges), then the total free \emph{and} bound charge on the object
is~$q/\kappa_{\mathrm{bg}}$.  In the numerical solution of
Eq.~\eqref{eq:matrixeq_surf} we should constrain the total charge of
each object to its exact value at every GMRES iteration. The first
reason for this is accuracy: errors in the net charge (monopole term)
can overwhelm relatively subtle dielectric effects. The second reason is
convergence rate: as demonstrated in Appendix~\ref{sec:exact_spectra},
the net charge on an object may correspond to an outlying eigenvalue of
the operator~$\mathcal{A}$; eliminating the corresponding eigenvector
component may significantly improve $\mathcal{A}$'s condition
number. The third reason is consistency: if a finite system is not kept
charge neutral, the operators $\mathcal{G}$ and~$\mathcal{A}$ become
ill-defined.

\subsection{Surface representation of free charge}
\label{sub:surface_representation}

In addition to fixing the net charge of each object to its exact
value during the GMRES iterations, there is another technique to
improve accuracy.  Typically, the dielectric object and
its free charge distribution are rigid. In this case, we care only about
the electric field that the free charge produces externally. Thus, we
may replace any distribution of internal free charge with an
\emph{equivalent} free charge distribution at the object
surface~\cite{jackson99}. If, instead, internal free charge were
present, there would be expected (partial) cancellations between the
internal charge and the bound surface charge. Small inaccuracies in the
Ewald solver would lead to inexact cancellation, a spurious monopole
moment, and potentially large numerical error. We avoid such
cancellation errors via the surface representation of free charge.

For each charged object, we may calculate the equivalent free surface
charge as follows. Consider a virtual system containing the internal
free charge, the object medium replaced by vacuum
($\kappa_{\mathrm{obj}}=1$), and the background medium replaced by a
conductor ($\kappa_{\mathrm{bg}}\rightarrow\infty$), in which the
``virtual'' electric field is zero. We use our dielectric method to
calculate the bound surface charge for this virtual system [with net
charge fixed to zero by Eq.~\eqref{eq:netcharge}].  By the principle of
superposition, the desired free surface charge distribution is then the
negative of the calculated virtual bound charge.

\subsection{Bound charge initialization}
\label{sub:bound_init}

In a molecular dynamics simulation, dielectric objects move only a small
amount during each time step. The bound charge~$\sb(t-\Delta t)$ that
was calculated at the previous time step may be used as the initial
guess for $\sb(t)$ at the current time step. In our study of interacting
dielectric spheres\cite{barros13a} we observed that this optimization
reduced the required GMRES iterations per time step from about 4 to~3
when the accuracy target was~$10^{-4}$.

\subsection{Direct residual}
\label{sub:direct_residual}

We save a call to the Ewald solver by avoiding the explicit calculation
of $b$. Instead, we compute the residual as
\begin{align}
  r^{(m)} &= b-\mathcal{A}x^{(m)} \nonumber \\
  & = \sf-\bar{\kappa}(\sf+x^{(m)}) -
  \epsilon_{0}\Delta\kappa\mathbf{E}^{(m)}\cdot\hat{n} \;.
\end{align}
Here we reuse $\mathbf{E}^{(m)}$ (the electric field due to both free
charge $\sf$ and estimated bound charge~$x^{(m)}$), which GMRES already
calculated to construct the Krylov space.  We also replace
Eq.~\eqref{eq:conv_criterion} with a convergence criterion that is
independent of~$b$, $||r^{(m)}||<10^{-4}|x^{(m)}|c$. We select $c$ to be
a ``typical'' dielectric constant. In a system containing only two types
of dielectric media, we choose $c=\bar{\kappa}$, the mean dielectric
constant.

\subsection{Energy calculation}

Equation~\eqref{eq:energybound} suggests calculating the energy in two
steps: (1) generate the potential $\psi = \mathcal{G}(\rf+\rb) /
\epsilon_{0}$ due to free and bound charge, and (2) sum the energy
contributions at the locations of free charge, $U = (1/2) \int \rf \psi
\mathd\mathbf{r}$.  The surface patch corrections described in
Sec.~\ref{sub:patch_corr} naturally extend to the calculation of the
potential $\psi_{ij}$ generated by surface patch~$j$ and evaluated at
surface patch~$i$.

Most electrostatics software packages do \emph{not} provide a procedure
to calculate $\psi$. However, these packages can still be used to
calculate the dielectric energy efficiently. Our trick is to express the
energy as
\begin{equation}
  U = \frac{1}{2} (\tuelec[\rf] - \tuelec[\rb] + \tuelec[\rf+\rb]) \;,
  \label{eq:uvac_tilde}
\end{equation}
where $\tuelec[\tilde \rho]$ represents the energy of the electric field generated by
the charge density~$\tilde \rho(\mathbf{r})$ alone,
\begin{equation}
  \tuelec[\tilde \rho] = \frac{1}{2\epsilon_{0}}\int_V
  \tilde \rho \mathcal{G} \tilde \rho \mathd\mathbf{r} \;.
\end{equation}
In particular, by Eqs.~\eqref{eq:gdef}, \eqref{eq:edef}, and~\eqref{eq:uvac},
\begin{equation}
  \uelec = \tuelec[\rf+\rb]
  \label{eq:uvac_explicit}
\end{equation}
is the bare electric field energy for the physical system.

Equation~\eqref{eq:uvac_tilde} states that we can calculate the full
dielectric energy using three separate calls to an Ewald solver. In
typical molecular dynamics applications, the energy is sampled at only a
small fraction of the time steps, and the cost of the two extra Ewald
evaluations is negligible.

\section{Comparison with previous numerical methods}
\label{sec:comparison}

\subsection{Richardson Iteration}
\label{sec:richardson}

Many existing simulation methods effectively calculate the bound charge
by Richardson iteration~\cite{Richardson10}, motivating us to consider
this case in detail.

The method proposed in Ref.~\onlinecite{levitt78} is perhaps the
earliest, and it iteratively calculates the surface bound charge
$\sigma_b = x$ via
\begin{equation}
  x^{(m+1)} =
  -\epsilon_{0} 
  \frac{\Delta\kappa}{\overline{\kappa}}\mathbf{E}^{(m)}\cdot\hat{n} 
  \;,
  \label{eq:levitt}
\end{equation}
where $\mathbf{E}^{(m)}$ is the electric field generated by both
non-surface charges and the surface bound charge $x^{(m)}$ from the previous iteration. Zero
free surface charge, $\sf=0$, is assumed. In the operator notation of
Eqs.~\eqref{eq:adef_surf} and~\eqref{eq:bdef_surf}, the recurrence
becomes
\begin{equation}
  x^{(m+1)}=x^{(m)}+\gamma(b-\mathcal{A}x^{(m)}) \;,
  \label{eq:richardson}
\end{equation}
where $\gamma^{-1}=\bar{\kappa}$ is the mean
dielectric constant at the surface. This numerical scheme, Richardson
iteration, is readily analyzed for
arbitrary~$\gamma$~\cite{Young89}. After some algebra, we express the
residual $r^{(m)} = b-\mathcal{A}x^{(m)}$ as a linear recurrence,
\begin{equation}
  r^{(m+1)}=(1-\gamma\mathcal{A})r^{(m)} \;.
\end{equation}
To solve this recurrence, we work in the basis of eigenvectors
$\{v_{\lambda}\}$ of the operator~$\mathcal{A}$. The residual vectors
become $r^{(m)}=\sum_\lambda r^{(m)}_{\lambda} v_{\lambda}$
and we obtain the solution
\begin{equation}
r^{(m+1)}_{\lambda} =
  (1-\gamma\lambda)r^{(m)}_{\lambda} = (1-\gamma\lambda)^{m+1}r^{(0)}_{\lambda}
  \;.
  \label{eq:richardsonerror}
\end{equation}
The consequence is that $r^{(m)}$ converges to zero if
$|1-\gamma\lambda|<1$ is satisfied for each eigenvalue
$\lambda$. Clearly $\gamma$ should be selected
according to the spectra of $\mathcal{A}$.

Somewhat remarkably, the implicit choice of Ref.~\onlinecite{levitt78},
$\gamma=2/(\kappa_{\mathrm{min}}+\kappa_{\mathrm{max}})$, leads to a
convergent scheme.  The eigenvalue bounds
$1\leq\kappa_{\min}\leq\lambda\leq\kappa_{\max}$ of
Eq.~\eqref{eq:eigenbounds} imply
\begin{equation}
  |1-\gamma\lambda| \leq 
  \frac{\kappa_{\mathrm{max}}-\kappa_{\mathrm{min}}}
  {\kappa_{\mathrm{max}}+\kappa_{\mathrm{min}}} < 1 \;.
\end{equation}
Although this scheme is consistent, other iterative solution methods
such as GMRES and BiCGSTAB are preferable for their much faster
convergence~\cite{Saad03}.

In the method of Ref.~\onlinecite{tyagi10}, Eq.~\eqref{eq:levitt} is
generalized to
\begin{equation}
  x^{(m+1)} = -\omega\epsilon_{0} \frac{\Delta\kappa}{\overline{\kappa}}
  \mathbf{E}^{(m)}\cdot\hat{n}+(1-\omega)x^{(m)} \;,
\end{equation}
for tunable~$\omega$. This scheme again corresponds to Richardson iteration,
Eq.~\eqref{eq:richardson}, now with step size $\gamma=\omega /
\bar{\kappa}$.

In a na{\"i}ve implementation, each Richardson iteration requires
$\mathcal O(n^2)$ operations to determine the electric field $\mathbf
E^{(m)}$ at all $n$ surface patches. As discussed in
Sec.~\ref{sec:matrix_vec}, this cost can be reduced to $\mathcal O(n \ln
n)$ or $\mathcal O(n)$ with a fast Ewald solver.

\subsection{Variational approaches}
\label{sec:variational}

In our review of linear dielectrics, Sec.~\ref{sec:review}, we
introduced the equilibrium polarization field as the one minimizing the
(free) energy functional $U = \uelec + \upol$. This
variational formulation of dielectrics can be used as the basis of
numerical methods~\cite{marchi01,levy05,maggs06,rottler09}, at the cost
of working with the bulk polarization (rather than just the surface
bound charge).  As we have seen, numerical efficiency is much improved
by posing the dielectric problem in terms of bound charge restricted to
the dielectric interfaces. Using an alternate variational formulation of
the dielectric problem~\cite{jackson99}, the authors of
Ref.~\onlinecite{allen01} determine the bound charge as the distribution
that minimizes a given functional. Subsequently, a similar functional
was found that, when minimized, corresponds to the
energy~\cite{jadhao12}. This latter approach enables Car--Parrinello
type molecular dynamics simulation.  Here, we analyze the computational
efficiency of numerical methods to calculate the bound charge based upon
these variational formulations.

The electrostatic energy may be expressed as the extremum of the functional~\cite{jadhao12}
\begin{equation}
  \mathcal{U}[\mathbf{P},\rb,\psi] =
  U-\int_V\psi(\br) [\rb(\br)
  +\nabla\cdot\mathbf{P}(\br)]\mathd\mathbf{r} \;.
  \label{eq:energy_functional}
\end{equation}
The Lagrange multiplier $\psi(\mathbf{r})$ in
Eq.~\eqref{eq:energy_functional} enforces the physical constraint
$\nabla\cdot\mathbf{P}=-\rb$. By Eqs.~\eqref{eq:u},~\eqref{eq:uvac_explicit} and~\eqref{eq:upol} the  electrostatic energy~$U = \uelec+\upol$ has the functional form
\begin{equation}
      U =\frac{1}{2\epsilon_{0}}
      \int_V \left[(\rf+\rb)\mathcal{G}(\rf+\rb)
        +\frac{\mathbf{P}^{2}}{\kappa-1}\right]\mathd \mathbf{r} \;.
\end{equation}
From $\mathcal U$, we wish to construct new functionals that are independent of $\mathbf P$ and $\psi$, and that are still extremized by the physical bound charge $\rb$. We extremize $\mathcal U$ with respect to $\rb$ and $\mathbf P$, and obtain
\begin{align}
  \psi       & = \mathcal{G}(\rf+\rb)/\epsilon_{0} \;,
  \label{eq:psi_var} \\
  \mathbf{P} & = -(\kappa-1)\nabla\mathcal{G}(\rf+\rb) \;.
  \label{eq:p_var}
\end{align}
Substitution of Eqs.\ \eqref{eq:psi_var} and~\eqref{eq:p_var} into
$\mathcal{U}$ yields the negative of the functional considered
in Ref.~\onlinecite{allen01},
\begin{equation}
  \mathcal{I}[\rb] = \frac{1}{2}\int_V
  \left[\rf\mathcal{G}(\rf+R_{b}[\rb])-\rb\mathcal{G}(\rb-R_{b}[\rb])\right]
  \mathd\mathbf{r} \;,
\end{equation}
where
\begin{equation}
  R_{b} = \nabla\cdot(\kappa-1)\nabla\mathcal{G}(\rf+\rb)
  = (1-\mathcal{A})(\rf+\rb) \;.
\end{equation}
We may also extremize $\mathcal U$ with respect to $\psi$, in which case $\rho_b = R_b[\rho_b]$.
Partial substitution then yields the alternate functional introduced in Ref.~\onlinecite{jadhao12},
\begin{equation}
  \mathcal{J}[\rb] = \frac{1}{2}\int_V
  \left[\rf\mathcal{G}(\rf+R_{b}[\rho_b]) -
    R_{b}[\rho_b]\mathcal{G}(\rb-R_{b}[\rho_b])\right]
  \mathd\mathbf{r} \;,
\end{equation}
The variation of $\mathcal I[\rb]$ and $\mathcal J[\rb]$ is readily calculated using the
identities
\begin{align}
  \int_V f\mathcal{G}\frac{\delta\rb}{\delta\rb}\mathd\mathbf{r}
  & = \mathcal{G}f \;,\\
  \int_V f\mathcal{G}\frac{\delta R_{b}}{\delta\rb}\mathd\mathbf{r}
  & = \mathcal{G}(1-\mathcal{A})f \;,
\end{align}
valid for any test function~$f(\mathbf{r})$. Extremization then yields
\begin{align}
  \frac{\delta\mathcal{I}}{\delta\rb}
  & = \mathcal{G}(b-\mathcal{A}\rb) = 0\;,\\
  \frac{\delta\mathcal{J}}{\delta\rb}
  & = \mathcal{G}(1-\mathcal{A})(b-\mathcal{A}\rb) = 0\;,
\end{align}
which are uniquely satisfied when $\mathcal{A}\rb=b$
[Eq.~\eqref{eq:matrixeq}], thus giving the correct bound charge.

Reference~\onlinecite{allen01} calculates $x=\rb$ by a steepest-ascent
procedure,
\begin{equation}
  x^{(m+1)} = x^{(m)}+\mathcal{\gamma}\frac{\delta\mathcal{I}}{\delta x^{(m)}}
  = x^{(m)}+\gamma\mathcal{G}(b-\mathcal{A}x^{(m)}) \;,
  \label{eq:hansenscheme}
\end{equation}
where $\gamma$ is a step size parameter. We recognize this variational
scheme as Richardson iteration, Eq.~\eqref{eq:richardson},
preconditioned by the positive definite operator~$\mathcal{G}$. By
Eq.~\eqref{eq:richardsonerror}, the convergence rate of Richardson
iteration is controlled by the ratio of extremal eigenvalues,
$\cnum = |\lambda_{\max}/\lambda_{\min}|$, of the relevant operator---in this
case~$\mathcal{GA}$.

We demonstrated in Sec.~\ref{sec:a_properties} that $\mathcal{A}$ is
well conditioned. In contrast, the eigenvalues of
$\mathcal{G}\mathcal{A}$ are unbounded in the continuum limit of small
patches---the operator has \emph{infinite} condition number. We use
simple scaling to compare the spectra of $\mathcal{A}$ and
$\mathcal{GA}$. The operator $\mathcal{A}$ is dimensionless and its
eigenvalues are independent of length scale. Since $\mathcal{G}$ is
inverse to~$\nabla^{2}$, it has dimensions of length squared.  The
operator $\mathcal{GA}$ inherits these dimensions. Thus, eigenvectors of
$\mathcal{G}\mathcal{A}$ with characteristic frequency~$k$ have
eigenvalues that scale as~$k^{-2}$. In the continuum limit, arbitrarily
small eigenvalues are possible. As a concrete example, consider the
uniform dielectric system $\kappa(\mathbf{r})=\kappa_{\mathrm{bg}}$
where $\mathcal{A=\kappa_{\mathrm{bg}}}$. The eigenvectors of
$\mathcal{GA=\kappa_{\mathrm{bg}}G}$ are the Fourier modes
$\exp(i\mathbf{k}\cdot\mathbf{r})$ with eigenvalues
$\kappa_{\mathrm{bg}}|k|^{-2}$ ranging from $0$ to~$\infty$.

In practice, the largest $k$-vector is cut off by the inter-patch
distance length scale. Similarly, the smallest $k$-vector is set by the
scale of the largest dielectric objects. Nonetheless, $\mathcal{GA}$ is
unnecessarily ill-conditioned. Thus the scheme of
Eq.~\eqref{eq:hansenscheme} requires many iterations for the bound
charge to converge.

These scaling considerations also apply to variational methods based on
$\mathcal{J}[\rb]$. Unlike $\mathcal{I}[\rb]$, the
functional~$\mathcal{J}[\rb]$ may be interpreted as an effective energy
functional in the sense that
\begin{equation}
  \min_{\rb}\mathcal{J}[\rb] = U \;,
\end{equation}
where $U$ is the usual dielectric
energy. Reference~\onlinecite{jadhao12} applied Car--Parrinello
molecular dynamics to evolve $\rb$ along with ion positions according to
the Hamiltonian~$\mathcal{J}$~\cite{Car85}.  An artificially low
temperature was separately applied to the $\rb$ degrees of
freedom. Thus, $\rb$ was effectively solved by the overdamped dynamics,
\begin{equation}
  \frac{\partial\rb(\br)}{\partial t} =
  -\gamma\frac{\delta\mathcal{J}}{\delta\rb(\br)} \;,
\end{equation}
for which we recover Eq.~\eqref{eq:hansenscheme} but now with
$\mathcal{G}(1-\mathcal{A})\mathcal{A}$ as the relevant operator. As
before, dimensional analysis tells us that
$\mathcal{G}(1-\mathcal{A})\mathcal{A}$ is ill-conditioned, and that
$\rb$ will converge slowly. In practice, this means that a very small
Car--Parrinello molecular dynamics time step must be employed.

\subsection{Induced Charge Computation (ICC) method}
\label{sec:icc}

The ICC method~\cite{boda04} proposed to solve
Eq.~\eqref{eq:matrixeq_surf} by explicit construction of the matrix
inverse $\mathcal{A}^{-1}$.  Direct matrix inversion costs
$\mathcal{O}(n^{3})$ for $n$ surface patch elements.  Subsequently, the
bound charge $x = \sb$ may be found by dense matrix--vector
multiplication, $x=\mathcal{A}^{-1}b$, where $b$ is a function of the
evolving free charge. For static dielectric geometries, each evaluation
of $x$ then costs $\mathcal{O}(n^{2})$, which is much worse than
$\mathcal{O}(n)$ methods based upon fast Ewald solvers.  If the
dielectric geometry dynamically evolves, then \emph{repeated} matrix
inversion is required, for which the authors of Ref.~\onlinecite{boda04}
suggested GMRES as an alternative.

\subsection{GMRES with fast matrix--vector product}

In light of the drawbacks of previously proposed methods, namely the use
of inefficient iterative methods employing Richardson iteration
(Sec.~\ref{sec:richardson}), an ill-conditioned operator and hence poor
convergence rates of variational methods (Sec.~\ref{sec:variational}),
and inefficient matrix--vector multiplication (and moreover matrix
inversion) for the ICC method (Sec.~\ref{sec:icc}), we advocate
calculation of the bound charge by solving Eq.~\eqref{eq:matrixeq_surf}
via GMRES and a fast Ewald solver.  With this approach, the surface
bound charge converges to high accuracy in only a handful of GMRES
iterations, each requiring $\mathcal O(n)$ or $\mathcal O(n \ln n)$
operations, depending on the Ewald solver used.

During the preparation of this publication, it came to our attention
that our strategy was proposed already in Ref.~\onlinecite{Bharadwaj95},
which has been overlooked and underappreciated, as evidenced by the wide
array of subsequent methods proposals.  We note, however, that due to
the large number of patches typically introduced for each dielectric
object, the acceleration techniques introduced in Sections
\ref{sub:fix_net_charge}--\ref{sub:direct_residual} are still
instrumental in realizing dynamic simulations such as those of
Ref.~\onlinecite{barros13a}.

\section{Summary}

In this paper we have demonstrated a collection of techniques by which
\emph{dynamic} dielectric systems can be simulated efficiently. In
geometries with sharp dielectric boundaries, one solves a matrix
equation to obtain the surface bound charge, from which energy and
forces follow directly. Empirically, we find that the bound charge
converges to high accuracy after a handful of GMRES iterations. We
attribute this fast convergence to the compact spectrum of the relevant
operator~$\mathcal{A}$, whose properties we have analyzed in detail.
Each iteration of GMRES requires only a single calculation of the
electric field in vacuum, which can be performed with an Ewald solver at
a cost that scales nearly linearly in the number of surface patch
elements~$n$.

Compared to several previous methods, our approach (i)~converges
quickly, by using GMRES rather than Richardson
iteration~\cite{levitt78,hoyles98,tyagi10}, (ii)~avoids the
ill-conditioned matrix equations of variational
approaches~\cite{allen01,jadhao12}, (iii)~does not require explicit
construction of the matrix inverse~\cite{boda04}, and (iv)~evaluates
matrix--vector products very efficiently with a fast Ewald solver. A
side benefit of (iv) is that we properly treat periodic geometries common
in computational studies.  To illustrate the capabilities of our method,
we have performed the first large-scale simulation of dynamical
dielectric objects in Ref.~\onlinecite{barros13a}.

\begin{acknowledgments}
  This material is based upon work supported by the National Science
  Foundation under Grant Nos.\ DMR-1006430 and~DMR-1310211. We
  acknowledge computing time at the Quest high-performance computing
  facility at Northwestern University. K.B. also acknowledges support by
  the LANL/LDRD program under the auspices of the DOE NNSA, contract
  number DE-AC52-06NA25396.
\end{acknowledgments}

\appendix

\section{Dielectric forces}
\label{sec:forces}

We derive forces in dielectric systems as a sum of pairwise Coulomb-like
interactions between free and bound charges. We follow the approach
advocated in Ref.~\onlinecite{Landau93} and carried out in
Refs.~\onlinecite{Schwinger98,zangwill13}. That is, we derive the force
on a dielectric object as the energy derivative with respect to object
motion.

Our first task is to express the electric field and energy as a function
of $\rf(\mathbf{r})$ and~$\kappa(\mathbf{r})$ alone. Combining Eqs.\
\eqref{eq:newconstitute} and~\eqref{eq:adef} we obtain
\begin{equation}
  \epsilon_{0}\psi = \mathcal{G}(\rf+\rb) =
  (-\nabla\cdot\kappa\nabla)^{-1}\rf \;.
\end{equation}
The existence of the symmetric operator $(\nabla\cdot\kappa\nabla)^{-1}$
follows from the existence of the potential $\psi$. The electric field
$\mathbf{E}=-\nabla\psi$ immediately follows,
\begin{equation}
  \mathbf{E} =
  \frac{1}{\epsilon_{0}}\nabla(\nabla\cdot\kappa\nabla)^{-1}\rf \;.
  \label{eq:field_f}
\end{equation}
The energy in Eq.~\eqref{eq:energybound} becomes a nonlocal,
$\kappa(\mathbf{r})$-dependent sum of free charge pairs,
\begin{equation}
  U = -\frac{1}{2\epsilon_{0}} 
  \int\rf(\nabla\cdot\kappa\nabla)^{-1}\rf\mathd\mathbf{r}
  \;.
  \label{eq:u1}
\end{equation}

\subsection{Force on free charge}

The force density~$\mathbf{f}$ associated with displacement of the free
charge~$\rf$ at position~$\mathbf{r}$ in any
direction~$\hat{\mathbf{n}}$ is given by
\begin{equation}
  \hat{\mathbf{n}}\cdot\mathbf{f} =
  -\lim_{\varepsilon\rightarrow0}
  \frac{U[\rf+\varepsilon\rho_{d}]-U[\rf]}{\varepsilon}
  \;,
\end{equation}
where the displacement distribution is
\begin{equation}
\rho_{d}(\mathbf{r}') = \rf(\mathbf{r})
\frac{\delta(\mathbf{r}+\varepsilon\hat{\mathbf{n}} -\mathbf{r}') -
  \delta(\mathbf{r}-\mathbf{r}')}{\varepsilon} \;.
\end{equation}
We expand in powers of~$\varepsilon$, dropping
$\mathcal{O}(\varepsilon)$ terms,
\begin{align}
  \frac{U[\rf+\varepsilon\rho_{d}]-U[\rf]}{\varepsilon} & \approx\int\frac{\delta U}{\delta\rf(\mathbf{r}')}\rho_{d}(\mathbf{r}')\mathd\mathbf{r}' \nonumber \\
  & \approx\rf\hat{\mathbf{n}}\cdot\nabla\frac{\delta U}{\delta\rf} \;.
\end{align}
The equality becomes exact in the limit $\varepsilon\rightarrow0$,
\begin{equation}
  \mathbf{f} = -\rf\nabla\frac{\delta U}{\delta\rf} \;.
  \label{eq:f_charge_def}
\end{equation}
Using both Eq.~\eqref{eq:field_f} and Eq.~\eqref{eq:u1}, we evaluate
\begin{equation}
  \mathbf{f} =
  \frac{\rf}{\epsilon_{0}}\nabla(\nabla \cdot \kappa\nabla)^{-1}\rf =
  \rf\mathbf{E} \;.
  \label{eq:f_charge}
\end{equation}
The net force to move the charge~$\rf(\mathbf{r})$ in a region~$\Omega$
is
\begin{equation}
  \mathbf{F}_{\mathrm{charge}}=\int_{\Omega}\rf\mathbf{E}\mathd\mathbf{r}
  \;.
  \label{eq:forcefree}
\end{equation}
In particular, the force on a point charge $\rf(\mathbf{r}) =
q\delta(\mathbf{r}-\mathbf{r}_{i})$ is simply~$q\mathbf{E}$.

\subsection{Force on dielectric object}

Dielectric object motion affects the energy through changes
in~$\kappa(\mathbf{r})$.  If the background has fixed
$\kappa_{\mathrm{bg}}$, then dielectric object motion corresponds to a
displacement of $\kappa(\mathbf{r})-\kappa_{\mathrm{bg}}$ at
each~$\mathbf{r}$. Analogous to Eq.~\eqref{eq:f_charge_def}, the force
density for this displacement is
\begin{equation}
  \mathbf{f} = -(\kappa-\kappa_{\mathrm{bg}})
  \nabla\frac{\delta U}{\delta\kappa} \;.
\end{equation}
We will use the identity
\begin{equation}
  \frac{\delta}{\delta\kappa}\mathcal{B}^{-1} =
  -\mathcal{B}^{-1}\frac{\delta\mathcal{B}}{\delta\kappa}\mathcal{B}^{-1}
  \;,
\end{equation}
which is a consequence of the product rule,
\begin{equation}
  0 = \frac{\delta}{\delta\kappa}(\mathcal{B}\mathcal{B}^{-1}) =
  \frac{\delta\mathcal{B}}{\delta\kappa}\mathcal{B}^{-1} +
  \mathcal{B}\frac{\delta\mathcal{B}^{-1}}{\delta\kappa} \;.
\end{equation}
Note that the operators $\mathcal{B}^{-1}$
and~$\delta\mathcal{B}/\delta\kappa$ do not generally commute.

Taking $\mathcal{B}=\nabla\cdot\kappa\nabla$, the functional derivative
of the energy, Eq.~\eqref{eq:u1}, evaluates to
\begin{align}
  \frac{\delta U}{\delta\kappa} &
  = +\frac{1}{2\epsilon_{0}}
  \int_{\Omega}\rf\mathcal{B}^{-1}\frac{\delta\mathcal{B}}{\delta\kappa}
  \mathcal{B}^{-1}\rf \mathd\mathbf{r}
  \nonumber \\
  & = -\frac{\epsilon_{0}}{2}
  \int_{\Omega}\mathbf{E}\cdot\frac{\delta\kappa}{\delta\kappa}\mathbf{E}
  \mathd\mathbf{r} \nonumber \\
  & =-\frac{\epsilon_{0}}{2}\mathbf{E}^{2} \;.
\end{align}
The minus sign appears after integrating by parts.

In index notation, where repeated indices denote summation, the
$\alpha^{\mathrm{th}}$ component of the force per volume is
\begin{align}
  f_{\alpha} & = \frac{\epsilon_{0}}{2}(\kappa-\kappa_{\mathrm{bg}})
  \partial_{\alpha}E_{\beta}E_{\beta} \nonumber \\
  & =\epsilon_{0}(\kappa-\kappa_{\mathrm{bg}})E_{\beta}\partial_{\alpha}E_{\beta}
  \;.
\end{align}
The electric field is a gradient, $E_{\beta}=-\partial_{\beta}\psi$, so
it follows that $\partial_{\alpha}E_{\beta}=\partial_{\beta}E_{\alpha}$
and
\begin{align}
  f_{\alpha} &
  = \epsilon_{0}(\kappa-\kappa_{\mathrm{bg}})E_{\beta}\partial_{\beta}E_{\alpha}
  \nonumber \\
  &
  =\epsilon_{0}\partial_{\beta}[(\kappa-\kappa_{\mathrm{bg}})E_{\beta}E_{\alpha}]
  -\epsilon_{0}[\partial_{\beta}(\kappa-\kappa_{\mathrm{bg}})E_{\beta}]E_{\alpha}
  \;.
\end{align}
Equivalently,
\begin{equation}
  \mathbf{f} =
  \epsilon_{0}\nabla\cdot
  [(\kappa-\kappa_{\mathrm{bg}})\mathbf{E}\otimes\mathbf{E}]
  - \epsilon_{0}[\nabla\cdot(\kappa-\kappa_{\mathrm{bg}})\mathbf{E}]
  \mathbf{E} \;.
\end{equation}
From Eqs.\ \eqref{eq:div_e} and~\eqref{eq:div_d} we have
\begin{align}
  \nabla\cdot(\kappa-\kappa_{\mathrm{bg}})\mathbf{E}
  & = \nabla\cdot\kappa\mathbf{E}-\kappa_{\mathrm{bg}}\nabla\cdot\mathbf{E}
  \nonumber \\
  & = \frac{1}{\epsilon_{0}}[\rf-\kappa_{\mathrm{bg}}(\rf+\rb)] \;,
\end{align}
yielding 
\begin{equation}
  \mathbf{f} = \epsilon_{0}\nabla\cdot[(\kappa-\kappa_{\mathrm{bg}})
  \mathbf{E}\otimes\mathbf{E}] + [\kappa_{\mathrm{bg}}(\rf+\rb)-\rf]\mathbf{E}
  \;.
\end{equation}

The net dielectric force, $\mathbf{F}_{\mathrm{diel}} =
\int_{\Omega}\mathbf{f}\mathd\mathbf{r}$, is an integral over a region
$\Omega$ enclosing the object and its surface. After applying Gauss's
theorem, the total force separates into a boundary term $\epsilon_{0}
\int_{\partial\Omega} (\kappa-\kappa_{\mathrm{bg}})
(\hat{n}\cdot\mathbf{E})\mathbf{E}\mathd\mathbf{s}$ and a bulk term
$\int_{\Omega}[\kappa_{\mathrm{bg}}(\rf+\rb)-\rf]\mathbf{E}\mathd\mathbf{r}$.
The boundary term is zero because, by construction, the integral is
evaluated where $\kappa(\mathbf{r})=\kappa_{\mathrm{bg}}$. The net
dielectric force on the object becomes
\begin{equation}
  \mathbf{F}_{\mathrm{diel}} =
  \int_{\Omega}[\kappa_{\mathrm{bg}}(\rf+\rb)-\rf] \mathbf{E} 
  \mathd\mathbf{r} \;,
\end{equation}
where $\rf$ has been treated as fixed.

Typically, free charge moves rigidly with the object, so we should also
include its force, Eq.~\eqref{eq:forcefree}. The \emph{total} force on
the dielectric object is then
\begin{equation}
  \mathbf{F} = \mathbf{F}_{\mathrm{charge}} + \mathbf{F}_{\mathrm{diel}}
  = \kappa_{\mathrm{bg}}\int_{\Omega}(\rf+\rb)\mathbf{E}\mathd\mathbf{r} \;.
\end{equation}
As a consistency check, note that in the special case where
$\kappa(\mathbf{r})=\kappa_{\mathrm{bg}}$ is constant, we have
$(\rf+\rb)=\rf/\kappa_{\mathrm{bg}}$ and the dielectric force is zero,
$\mathbf{F}_{\mathrm{diel}}=\mathbf 0$.

\section{Exact spectra for simple geometries}
\label{sec:exact_spectra}

For certain dielectric geometries the entire spectrum of $\mathcal{A}$
can be determined. The key observation is that the eigenvectors of
$\mathcal{A}$ coincide with the solutions of the Laplace equation
$\nabla^{2}\psi=0$ in non-Cartesian coordinates, when those solutions
are separable in the normal component. This solution technique applies
to the dielectric sphere, cylinder, and slab. In these geometries,
$\mathcal{A}$ becomes a symmetric operator.

We seek eigenvectors~$\rho$ and eigenvalues~$\lambda$ that satisfy
$\mathcal{A}\rho = \lambda\rho$.  We work with surface charge
density~$\sigma$, for which Eq.~\eqref{eq:adef_surf} states
\begin{equation}
  \mathcal{A}\sigma =
  \bar{\kappa}\sigma+\Delta\kappa\mathbf{E}\cdot\hat{n} = \lambda\sigma
  \;,
  \label{eq:asurface}
\end{equation}
with $\mathbf{E}=(\mathbf{E}_{\mathrm{out}}+\mathbf{E}_{\mathrm{in}})/2$
the electric field \emph{at} the surface. We also have
$\Delta\kappa=\kappa_{\mathrm{out}}-\kappa_{\mathrm{in}}$, and
$\bar{\kappa}=(\kappa_{\mathrm{out}}+\kappa_{\mathrm{in}})/2$.  In this
section we use dimensionless units where $\epsilon_{0} = 1$.

\subsubsection{Sphere}

Consider a single spherical object of radius~$R$. We work in spherical
coordinates $(r,\theta,\phi)$. The operator~$\mathcal{A}$ is fixed upon
the specification
\begin{equation}
  \kappa(r)=\left\{
    \begin{array}{ll}
      \kappa_{\mathrm{in}} & \mathrm{if}\,\, r<R\\
      \kappa_{\mathrm{out}} & \mathrm{if}\,\, r>R
    \end{array}\right. \;.
\end{equation}
The spherical harmonics~$Y_{lm}(\theta,\phi)$ form an orthogonal basis
for the surface of the sphere. We will demonstrate that the spherical
harmonics are in fact the eigenvectors of~$\mathcal{A}$.  In
anticipation of this result, consider the surface charge distribution,
\begin{equation}
  \sigma(\theta,\phi)=Y_{lm}(\theta,\phi) \;.
\end{equation}
The electrostatic potential due to $\sigma$ is
\begin{equation}
  \psi(r,\theta,\phi) = \left\{
    \begin{array}{ll}
      \psi_{1}=ar^{l}Y_{lm}    & \mathrm{if}\,\, r<R\\
      \psi_{2}=br^{-l-1}Y_{lm} & \mathrm{if}\,\, r>R
    \end{array}\right. \;,
\end{equation}
where
\begin{align}
  a & = \frac{\sigma_{0}}{2l+1}R^{-l+1} \;, \\
  b & = \frac{\sigma_{0}}{2l+1}R^{l+2}  \;.
\end{align}
As required, $\psi$ satisfies the Laplace equation $\nabla^{2}\psi=0$
for $r\neq R$, and obeys appropriate boundary conditions at $r=R$:
\begin{align}
  \psi_{2}-\psi_{1}|_{r=R} & = 0  \;, \\
  \partial_{r}\psi_{2}-\partial_{r}\psi_{1}|_{r=R} & = -\sigma \;.
\end{align}
The electric field projected onto the surface normal is
\begin{equation}
  \hat{r}\cdot\mathbf{E} =
  -\frac{(\partial_{r}\psi_{1}+\partial_{r}\psi_{2})}{2} =
  -\frac{\sigma}{2(2l+1)}[-(l+1)+l] \;.
\end{equation}
Comparison with Eq.~\eqref{eq:asurface} confirms that $\sigma$ is indeed
an eigenvector,
\begin{align}
  \mathcal{A\sigma} = \lambda\sigma
  & = \bar{\kappa}\sigma+\Delta\kappa\mathbf{E}\cdot\hat{r}\nonumber \\
  & =\left(\bar{\kappa}+\frac{\Delta\kappa}{2(2l+1)}\right)\sigma \;.
\end{align}
Expanding $\bar{\kappa}=(\kappa_{\mathrm{out}}+\kappa_{\mathrm{in}})/2$
and $\Delta\kappa=\kappa_{\mathrm{out}}-\kappa_{\mathrm{in}}$ we get
\begin{equation}
  \lambda
  =\left\{ \kappa_{\mathrm{out}},
    \left(\frac{2}{3}\kappa_{\mathrm{out}}+\frac{1}{6}\kappa_{\mathrm{in}}\right),
    \ldots, \left( \frac{1}{2}\kappa_{\mathrm{out}}
      +\frac{1}{2}\kappa_{\mathrm{in}} \right) \right\} \;.
  \label{eq:lambda_sphere}
\end{equation}
The eigenvalue~$\lambda=\kappa_{\mathrm{out}}$ corresponds to the
eigenvector of uniform surface charge, $Y_{l=0,m=0}$. In a numerical
implementation, we constrain the net surface charge to its exact value
as described in Sec.~\ref{sub:fix_net_charge}, effectively eliminating
this eigenvector from the space. The eigenvalue
$\lambda=\kappa_{\mathrm{out}}$ should therefore be ignored.

The ratio $\cnum = \lambda_{\max}/\lambda_{\min}$ is greatest when
$\kappa_{\mathrm{out}}\gg\kappa_{\mathrm{in}}$ or
$\kappa_{\mathrm{in}}\gg\kappa_{\mathrm{out}}$, where $\cnum \approx
4/3$ or~$3$, respectively.

\subsubsection{Cylinder}

We now adopt cylindrical coordinates $(\rho,\theta,z)$ and consider a
dielectric cylinder,
\begin{equation}
  \kappa(\rho)=\left\{ 
    \begin{array}{ll}
      \kappa_{\mathrm{in}}  & \mathrm{if}\,\,\rho<R \\
      \kappa_{\mathrm{out}} & \mathrm{if}\,\,\rho>R
    \end{array}\right. \;.
\end{equation}
We will show that the eigenvectors of $\mathcal{A}$ take the form
\begin{equation}
\sigma=e^{ikz+i\nu\theta}
\end{equation}
for real wave number~$k$ and integer wave number~$\nu$. The functions
$\sigma(z,\theta)$ are the Fourier modes of the cylinder surface and
form a complete basis.

The electrostatic potential for~$\sigma$ is
\begin{equation}
\psi(\rho,\theta,z) = \left\{
  \begin{array}{ll}
    \psi_{1} & \mathrm{if}\,\,\rho<R\\
    \psi_{2} & \mathrm{if}\,\,\rho>R
  \end{array}\right. \;,
\end{equation}
where
\begin{align}
  \psi_{1} & =[aK_{\nu}(kR)]I_{\nu}(k\rho)e^{ikz+i\nu\theta} \;, \\
  \psi_{2} & =[aI_{\nu}(kR)]K_{\nu}(k\rho)e^{ikz+i\nu\theta} \;, \\
  a & =-\frac{1}{k}\left[I_{\nu}(kR)K_{\nu}'(kR) -
    I_{\nu}'(kR)K_{\nu}(kR)\right]{}^{-1} \;.
\end{align}
$I_{\nu}$ and $K_{\nu}$ are the modified Bessel functions of the first
and second kind, and primes denote derivatives:
$I_\nu'(x) = \mathd I(x)/\mathd x$ and $K_\nu'(x) = \mathd K(x)/\mathd x$.
As required, $\psi$ satisfies the Laplace equation $\nabla^{2}\psi=0$
for~$\rho\neq R$ and obeys appropriate boundary conditions at~$\rho=R$,
\begin{align}
  \psi_{2}-\psi_{1}|_{\rho=R} & = 0 \;, \\
  \partial_{\rho}\psi_{2}-\partial_{\rho}\psi_{1}|_{\rho=R} & = -\sigma \;.
\end{align}
The induced electric field projected onto the surface normal is
\begin{align}
    \mathbf{E}\cdot\hat{\rho}
    &
    =-\frac{(\partial_{\rho}\psi_{1}+\partial_{\rho}\psi_{2})|_{\rho=R}}{2}
    \nonumber \\ 
    & =\frac{1}{2}\left(\frac{1+C}{1-C}\right)\sigma \;,
\end{align}
where
\begin{equation}
  C(\nu,kR) = \frac{I_{\nu}'(kR)K_{\nu}(kR)}{I_{\nu}(kR)K_{\nu}'(kR)} \;.
\end{equation}
Comparison with Eq.~\eqref{eq:asurface} confirms that $\sigma$ is indeed
an eigenvector, with eigenvalue
\begin{equation}
  \lambda = \bar{\kappa}+\frac{\Delta\kappa}{2}\frac{1+C}{1-C} \;.
  \label{eq:lambda_cyl}
\end{equation}
The eigenvalues $\lambda$ are determined by the function $C(\nu,kR)$,
which satisfies $-1\leq C \leq 0$. The maximum of $C$ occurs at
low-frequency modes: $C\rightarrow0$ when $\nu=0$ and
$kR\rightarrow0$. Conversely, $C\rightarrow-1$ for high frequencies
$kR\rightarrow\infty$. The extreme eigenvalues follow immediately,
\begin{equation}
  \lambda\rightarrow\left\{
    \begin{array}{ll}
      \bar{\kappa}+\frac{1}{2}\Delta\kappa=\kappa_{\mathrm{out}}
      & \mathrm{if}\,\,(\nu=0,kR \rightarrow 0)\\
      \bar{\kappa}=\frac{1}{2}(\kappa_{\mathrm{out}}+\kappa_{\mathrm{in}})
      & \mathrm{if}\,\, kR\rightarrow\infty
    \end{array}\right. \;.
\end{equation}
The ratio $\cnum = \lambda_{\max}/\lambda_{\min}$ is greatest when
$\kappa_{\mathrm{in}}\gg\kappa_{\mathrm{out}}$, where $\cnum \approx
\kappa_{\mathrm{in}}/(2\kappa_{\mathrm{out}})$. In the limit
$\kappa_{\mathrm{out}}\gg\kappa_{\mathrm{in}}$ we find $\cnum \approx
2$.

If the length of the cylinder~$L$ is not too much greater than the
radius~$R$, then $\cnum$ can be reasonable even in the limit
$\kappa_{\mathrm{in}}\gg\kappa_{\mathrm{out}}$. For finite~$L$, we
ignore fringe effects and assume that the above analysis is
approximately correct with axial wave numbers taking discrete values
$k=\frac{2\pi}{L}\{0,1,\ldots\}$. As in the spherical case, the zeroth
mode represents a uniform charge distribution, and can be manually
removed from the vector space. If $L/R$ is not too large then
$C(\nu=0,kR\ll1)$ deviates significantly from~$0$, increasing the
associated eigenvalue. For example, if we choose $L/R=30$ and $k=2\pi/L$
then $C(0,kR)\approx-0.039$. Assuming
$\kappa_{\mathrm{in}}\gg\kappa_{\mathrm{out}}$, the smallest eigenvalue
is approximately $0.038\kappa_{\mathrm{in}}$, yielding $\cnum \approx
13.3$ (independent of the
ratio~$\kappa_{\mathrm{in}}/\kappa_{\mathrm{out}}$).

\subsubsection{Slab}

The final case to be considered is the slab, where we adopt cartesian
coordinates $(x,y,z)$ and choose
\begin{equation}
\kappa(\rho)=\left\{
  \begin{array}{ll}
    \kappa_{\mathrm{in}} & \mathrm{if}\,\,|z|<R\\
    \kappa_{\mathrm{out}} & \mathrm{if}\,\,|z|>R
  \end{array}\right. \;.
\end{equation}
The eigenvectors of $\mathcal{A}$ will be defined by their surface
densities on the two planes $z=\pm R$. There are two classes of
eigenvectors, symmetric and antisymmetric, represented as
\begin{equation}
  \sigma(x,y,R)=\pm\sigma(x,y,-R)=e^{ik_{x}x+ik_{y}y} \;.
\end{equation}
The eigenvectors are complete: an arbitrary distribution of charge on
\emph{both} planes can be represented in the basis of symmetric and
antisymmetric eigenvectors. The electrostatic potential in the symmetric
case is
\begin{equation}
\psi_{s}(x,y,z) = \left\{
  \begin{array}{ll}
    b_{s}\sigma e^{\gamma z} & \mathrm{if}\,\, z<-R\\
    a\sigma\cosh(\gamma z) & \mathrm{if}\,\,|z|<R\\
    b_{s}\sigma e^{-\gamma z} & \mathrm{if}\,\, z>+R
  \end{array}\right. \;,
\end{equation}
and for the antisymmetric case,
\begin{equation}
  \psi_{a}(x,y,z) = \left\{
    \begin{array}{ll}
      -b_{a}\sigma e^{\gamma z} & \mathrm{if}\,\, z<-R\\
      a\sigma\sinh(\gamma z)  & \mathrm{if}\,\,|z|<R\\
      b_{a}\sigma e^{-\gamma z} & \mathrm{if}\,\, z>+R
    \end{array}\right. \;,
\end{equation}
where 
\begin{align}
  a      & = \frac{e^{-\gamma R}}{\gamma}     \;, \\
  b_{s}   & = \frac{\cosh(\gamma R)}{\gamma} \;, \\
  b_{a}   & = \frac{\sinh(\gamma R)}{\gamma} \;, \\
  \gamma & = \sqrt{k_{x}^{2}+k_{y}^{2}} \;.
\end{align}
As required, $\psi_{s}$ and~$\psi_{a}$ satisfy the Laplace equation in
the bulk, and the usual boundary conditions at $z=\pm R$. The induced
electric fields at $z=R$, projected onto $\hat{z}$, are
\begin{equation}
  \hat{z}\cdot\mathbf{E}=\pm\frac{1}{2}\exp(-2\gamma R)\sigma \;,
\end{equation}
where $\pm$ refers to symmetric and antisymmetric eigenvectors,
respectively.  Comparison with Eq.~\eqref{eq:asurface} confirms that the
$\sigma$ (symmetric and antisymmetric) are indeed eigenvectors with
eigenvalues,
\begin{equation}
  \lambda=\bar{\kappa}\pm\frac{\Delta\kappa}{2}\exp(-2\gamma R) \;.
  \label{eq:lambda_slab}
\end{equation}
In the high-frequency limit ($k_{x}^{2}+k_{y}^{2}\rightarrow\infty$) the
constant $\gamma$ diverges and the eigenvalues tend to
$\lambda\rightarrow\kappa=(\kappa_{\mathrm{out}}+\kappa_{\mathrm{in}})/2$.
In the opposite limit, where the two planes each have nearly uniform
charge, the eigenvalues tend to
$\lambda\rightarrow\kappa_{\mathrm{out}}$ and~$\kappa_{\mathrm{in}}$ for
symmetric and antisymmetric cases, respectively. In these limits, $\cnum
= \kappa_{\max}/\kappa_{\min}$, realizing the worst-case behavior
allowed by the bounds of Eq.~\eqref{eq:eigenbounds}!

\section{Dielectric energies for simple geometries}
\label{sec:energies}

In Appendix~\ref{sec:exact_spectra} we studied the exact spectra of a
dielectric sphere, cylinder, and slab, and found that $\mathcal{A}$ is
generally well-conditioned, except for extreme dielectric contrasts
($\kappa_{\mathrm{in}}\ll\kappa_{\mathrm{out}}$ or
$\kappa_{\mathrm{out}}\ll\kappa_{\mathrm{out}}$) in the extended
cylinder or slab geometries. Here we demonstrate that, in geometries
where $\mathcal{A}$ remains well-conditioned, the energetics saturates
quickly as a function of the dielectric contrast.

The scaled energies of a point charge~$q$ interacting with dielectric
sphere, cylinder, and slab objects are~\cite{iversen98,cui06},
\begin{widetext}
\begin{align}
  \frac{U_{\mathrm{sphere}}}{u_{0}}   & =
  2\frac{d}{r_{0}}\sum_{n=0}^{\infty}
  \frac{(1-\tilde{\kappa})n(1+d/r_{0})^{-2(n+1)}}{(1+\tilde{\kappa})n+1} \;,
  \label{eq:usphere}\\
  \frac{U_{\mathrm{cylinder}}}{u_{0}} & = -\frac{8}{\pi}\frac{d}{r_{0}}
  \int_{0}^{\infty}\left(\frac{1}{2}a_{0}(u)+\sum_{n=1}^{\infty}a_{n}(u)\right)
  \mathd u; \,\,\,\,\,\, 
  a_{n}(u) = \frac{(1-\tilde{\kappa})K_{n}^{2}((1+d/r_{0})u)}%
  {\tilde{\kappa}\frac{K_{n}(u)}{I_{n}(u)} - 
    \frac{\partial_{u}K_{n}(u)}{\partial_{u}I_{n}(u)}}\;, \label{eq:ucylinder}\\
  \frac{U_{\mathrm{slab}}}{u_{0}} & =
  \frac{1-\tilde{\kappa}}{1+\tilde{\kappa}} - 
  \frac{4\tilde{\kappa}}{(1+\tilde{\kappa})^{2}}
  \sum_{n=1}^{\infty}
  \left(\frac{1-\tilde{\kappa}}{1+\tilde{\kappa}}\right)^{2n-1}
  \left(1+\frac{2n}{d/r_{0}}\right)^{-1} \;,
  \label{eq:uslab}
\end{align}
\end{widetext} where $r_{0}$ is the radius of the dielectric object (for
the slab, $r_{0}$ is half the thickness), $d$ is the distance between
the point charge and object surface, and $I_{n}$ and~$K_{n}$ are again
the modified Bessel functions. The dielectric constants control the
contrast $\tilde{\kappa}=\kappa_{\mathrm{obj}}/\kappa_{\mathrm{bg}}$,
Eq.~\eqref{eq:contrast}, and the reference energy scale,
\begin{equation}
  u_{0} = \frac{q^{2}}{16\pi\epsilon_{0}\kappa_{\mathrm{bg}}d} \;.
\end{equation}

In the limit that the point charge approaches the object surface, all
three geometries are effectively equivalent to a simple flat plane, and
the three energies converge to
\begin{equation}
  \frac{U_{\mathrm{plane}}}{u_{0}}
  = \frac{1-\tilde{\kappa}}{1+\tilde{\kappa}} 
  = -\tanh\left(\frac{\ln\tilde{\kappa}}{2}\right);\,\, 
  d\ll r_{0} \;.
  \label{eq:uplane}
\end{equation}
Saturation occurs quickly at the \emph{conducting limits} where
$\ln\tilde{\kappa}\rightarrow\pm\infty$.  At $\tilde{\kappa}=10^{\pm1}$
the energy~$U_{\mathrm{plane}}$ is within 20\% of its limiting values.

\begin{figure}
  \includegraphics{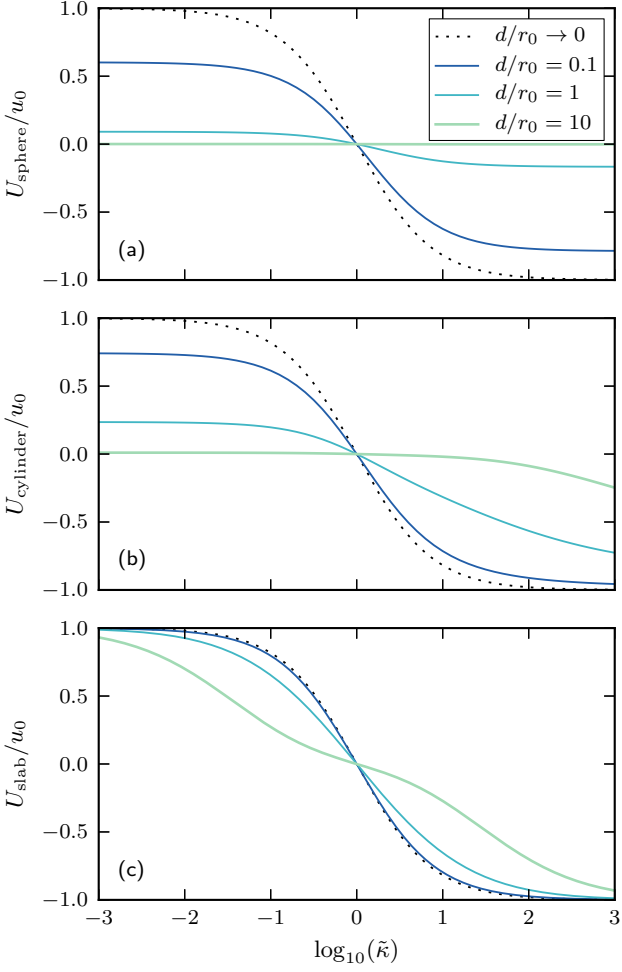}
  \caption{Scaled energies for a point charge at distance~$d$ from the
    surface of a (a)~dielectric sphere, (b)~cylinder, and (c)~slab. The
    dielectric contrast $\tilde{\kappa} =
    \kappa_{\mathrm{obj}}/\kappa_{\mathrm{bg}}$ controls the magnitude
    of dielectric effects. There are three different limiting behaviors
    when $d\rightarrow\infty$: (a)~$U_{\mathrm{sphere}}/u_{0}$ goes to
    zero, (b)~$U_{\mathrm{cylinder}}/u_{0}$ goes to zero except when
    $\ln\tilde{\kappa}\rightarrow+\infty$, where it goes to~$-1$,
    (c)~$U_{\mathrm{slab}}/u_{0}$ goes to $\mp 1$ when
    $\ln\tilde{\kappa}\rightarrow\pm\infty$.}
\label{fig:contrasts}
\end{figure}

In Fig.~\ref{fig:contrasts} the scaled energies are plotted as functions
of $\log_{10} \tilde{\kappa}$. When $d\ll r_{0}$, we recover the
asymptotic behavior in Eq.~\eqref{eq:uplane}. However, when $d\gg
r_{0}$, the sphere, cylinder, and slab geometries differ
markedly. The sphere energy decays like $d^{-4}$ when $d\gg r_{0}$, and
$U_{\mathrm{sphere}}/u_{0}$ goes to $0$ even when
$\ln\tilde{\kappa}\rightarrow\pm\infty$. The cylinder energy exhibits a
pronounced asymmetry: $U_{\mathrm{cylinder}}/u_{0}$ goes to $0$ when
$d\gg r_{0}$, \emph{except} when $\ln\tilde{\kappa}\rightarrow+\infty$,
where $U_{\mathrm{cylinder}}/u_{0}$ goes to $-1$. The slab energy is
antisymmetric in dielectric contrast,
$U_{\mathrm{slab}}(\tilde{\kappa})=-U_{\mathrm{slab}}(1/\tilde{\kappa})$. It
also responds most strongly, with a scaled energy
$U_{\mathrm{slab}}/u_{0}$ that goes to $\mp1$ in both conducting limits
$\ln\tilde{\kappa}\rightarrow\pm\infty$, independent of $d$.

The above energy scaling has an interesting connection to the spectrum
of $\mathcal A$. In Appendix~\ref{sec:exact_spectra} we solved the
exact spectrum of $\mathcal A$ for sphere, cylinder, and slab
geometries, and found that the ratio of extremal eigenvalues $\cnum$ is
large precisely when the dielectric interaction $U / u_0$ is abnormally large: the
cylinder when $\ln\tilde{\kappa}\rightarrow\infty$ and the slab when
$\ln\tilde{\kappa}\rightarrow\pm\infty$.

\bibliographystyle{apsrev4-1}
\bibliography{bibtex/journals,bibtex/colloids,bibtex/dielectrics,bibtex/electrolyte,bibtex/misc,bibtex/simu,bibtex/refs2}

\end{document}